\documentclass[12pt]{article}

\usepackage{epsf,epsfig,epstopdf}
\usepackage{graphics}

\usepackage{url}
\usepackage{amsmath,amssymb}
\usepackage{cite}
\usepackage{a4wide}
\numberwithin{equation}{section}
\newcommand{\prog}{{\sc Topixs}}

\begin{document}

\allowdisplaybreaks
\thispagestyle{empty}

\begin{flushright}
{\small
TUM-HEP-842/12\\
TTK-12-23\\
ITP-UU-12/21\\
SPIN-12/19\\
FR-PHENO-2012-010\\
SFB/CPP-12-34\\
1206.2454 [hep-ph]\\
July 27, 2012}
\end{flushright}

\begin{center}
\vspace{0.9\baselineskip}
\textbf{\Large\bf Inclusive top-pair production phenomenology 
with T{\normalsize OPIXS}
\\[0.2cm]}

\vspace{2\baselineskip}
{\sc M.~Beneke$^{a,b}$, P.~Falgari$^c$, S.~Klein$^b$, J.~Piclum$^{a,b}$, 
C.~Schwinn$^d$,\\ M.~Ubiali$^b$, F.~Yan$^b$}\\
\vspace{0.7cm}
{\sl ${}^a$Physik Department T31,\\
James-Franck-Stra\ss e, Technische Universit\"at M\"unchen,\\
D--85748 Garching, Germany\\
\vspace{0.3cm}
${}^b$Institut f\"ur Theoretische Teilchenphysik und 
Kosmologie,\\
RWTH Aachen University, D--52056 Aachen, Germany\\
\vspace{0.3cm}
${}^c$Institute for Theoretical Physics and Spinoza Institute,\\
Utrecht University, 3508 TD Utrecht, The Netherlands\\
\vspace{0.3cm}
${}^d$ Albert-Ludwigs Universit\"at Freiburg, 
Physikalisches Institut, \\
D-79104 Freiburg, Germany}

\vspace*{1cm}
\textbf{Abstract}\\
\vspace{1\baselineskip}
\parbox{0.9\textwidth}{ We discuss various aspects of 
  inclusive top-quark pair production based on \prog, a new, flexible
  program that computes the production cross section at the Tevatron
  and LHC at next-to-next-to-leading logarithmic accuracy in soft and
  Coulomb resummation, including bound-state effects and the complete
  next-to-next-to-leading order result in the $q\bar q$ channel, which
  has recently become available. We present the calculation of the
  top-pair cross section in $pp$ collisions at $8\,$TeV centre-of-mass
  energy, as well as the cross sections for hypothetical heavy 
  quarks in extensions of the standard model. The dependence on the
  parton distribution input is studied.  Further we investigate the
  impact of LHC top cross section measurements at $\sqrt{s}=7\,$TeV on
  global fits of the gluon distribution using the NNPDF re-weighting
  method.  }
\end{center}

\newpage
\setcounter{page}{1}
%%%%%%%%%%%%%%%%%%%%%%%%%%%%%%%%%%%%%%%%%%%%%%%%%%%%%%%%%%%%%%%%

\section{Introduction}

The measurement of the 
inclusive top-quark pair production cross section is currently 
of great interest at the Large Hadron Collider (LHC). 
It provides constraints on models that explain the anomalous 
forward-backward asymmetry by new physics; measures the top-quark 
mass cleanly, though not very precisely; 
is sensitive to the gluon distribution in the 
proton in a momentum-fraction region 
that is not well constrained by other measurements; and it serves as 
a template for searches for unknown particles with missing transverse 
momentum. For all these reasons a precise cross section calculation is 
desirable, and, in fact, the top-quark pair production cross section 
is by now the most precisely predicted purely hadronic high-energy 
cross section. 

In \cite{Beneke:2011mq} some of us presented a calculation of the 
top-quark pair production cross section that included the 
threshold approximation to the  
next-to-next-to-leading-order (NNLO) partonic cross 
sections \cite{Beneke:2009ye}, the joint resummation of soft and 
Coulomb corrections to all orders with next-to-next-to-leading 
logarithmic (NNLL) accuracy, and bound-state effects due to 
Coulomb attraction. The calculation 
of \cite{Beneke:2011mq} therefore constitutes the most complete 
implementation of the inclusive production cross section. 
With the present paper we release the first version of the user-friendly 
public program \prog \mbox{ }(``TOP-pair Inclusive X[cross] Section''), which 
incorporates the calculation of \cite{Beneke:2011mq} and an important 
update as described below. The functionality of the program is 
described in the appendix of this paper. The program, together with 
a detailed manual, can be obtained from the URL 
\begin{center}
\verb+http://users.ph.tum.de/t31software/topixs/+
\end{center}
which will be continuously updated.  Other publicly available
implementations beyond NLO exist
\cite{Aliev:2010zk,Cacciari:2011hy,Czakon:2011xx} and agree at the
few-percent level, but do not include soft-gluon resummation (HATHOR)
and Coulomb resummation (HATHOR, top++) to all orders, while the
program provided with~\cite{Ahrens:2011px} obtains the inclusive cross
section from resummed or approximated differential distributions.
Further approximate NNLO results have been obtained
in~\cite{Kidonakis:2010dk,Brodsky:2012sz}.

In the following sections of this work, we consider various topics 
extending our previous results~\cite{Beneke:2011mq}, 
which are motivated by recent theoretical 
developments and the anticipated data of the 2012 LHC run:
\begin{itemize}
\item We include the (almost) exact NNLO $q\bar q$ partonic 
cross section, which has recently become 
available~\cite{Baernreuther:2012ws}. This decreases the cross 
section in $p\bar p$ collisions at the Tevatron by about 1\%, 
together with its theoretical uncertainty, 
but has a negligible effect on $pp$ collisions (LHC).
\item We provide results for $\sqrt{s}=8\,$TeV, since meanwhile the 
LHC centre-of-mass energy has been increased to this value.
\item In \cite{Beneke:2011mq} we exclusively used the 
MSTW2008 parton distribution functions (PDFs)~\cite{Martin:2009iq}. 
Here we compare the predictions for different PDF sets, since, given 
the accuracy of the partonic calculation, the PDF error is now 
the most important single theoretical uncertainty. Results 
for different PDFs based on the program HATHOR \cite{Aliev:2010zk} 
can also be found in \cite{Alekhin:2012ig}.
\item It is very likely that a 
perturbative sequential fourth generation will be excluded (or 
discovered) in the 2012 LHC run. Heavy vector-like fermions with 
Standard Model (SM) QCD interactions may exist in other models. 
We therefore present 
production cross sections for a hypothetical heavy ``top'' 
quark. 
\item The possibility to constrain the gluon distribution in the 
proton through the top production cross section has been 
mentioned in \cite{Nadolsky:2008zw,Watt:2011kp}. 
With precise theoretical calculations and LHC data at 
$\sqrt{s}=7\,$TeV, this can now be investigated in practice. 
We quantify the impact of the existing data on global fits 
of the gluon distribution using the NNPDF re-weighting method presented 
in~\cite{Ball:2010gb,Ball:2011gg}, including the case when
the jet data is left out from the ``global'' fit.
\end{itemize}
The reader interested in the theoretical framework of soft-gluon 
and Coulomb resummation is referred 
to \cite{Beneke:2011mq} and \cite{Beneke:2009rj,Beneke:2010da}.

\section{Updated cross-sections}
\label{sec:nnlo}

\subsection{Theoretical framework for NNLO+NNLL}
The total partonic top-pair production 
cross section depends on the top-quark mass and 
the kinematic variable $\beta=\sqrt{1-4 m_t^2/\hat s}$.
In \cite{Beneke:2011mq} singular terms as $\beta\to 0$ were 
summed to all orders at NNLL accuracy by representing the 
cross section in the form~\cite{Beneke:2010da} 
\begin{equation}
\label{eq:fact}
  \hat\sigma_{pp'}^{\text{NNLL}}(\hat s,\mu)
= \sum_{R={\bf 1},{\bf 8}}H^{R}_{pp'}(m_t,\mu)
\;\int d \omega\;
J_{R}(E-\frac{\omega}{2})\,
W^{R}(\omega,\mu)\, ,
\end{equation}
where $E=\sqrt{\hat s}-2 m_t$ is the energy relative to the production
threshold. The soft functions $W_R$ sum logarithms of $\beta$ from
soft-gluon emission. The potential function $J_R$ sums Coulomb-gluon
exchange related to the attractive or repulsive Coulomb force in the
colour-singlet and octet channels, respectively. Logarithms of $\beta$
from non-relativistic physics are currently included at NNLO in
$J_R$. The hard functions $H_R$ account for the short-distance
production process of the $t\bar t$ pair.  The momentum-space approach
to threshold resummation~\cite{Becher:2006nr,Becher:2007ty} is
employed, where renormalization-group equations evolve the hard
functions from a hard scale $\mu_h$ to the factorization scale $\mu_f$
and the soft function from the soft scale $\mu_s\sim m_t\beta^2$ to
$\mu_f$. As default the values $\mu_h=2m_t$, $\mu_f=m_t$ and a
``running'' soft scale $\mu_s=2m_t\, \text{max}[\beta^2,
\beta_{\text{cut}}^2]$ is used.  An alternative method with a fixed
soft scale, and the determination of $\beta_{\text{cut}}$ are
discussed in detail in~\cite{Beneke:2011mq}.

Since (\ref{eq:fact}) is strictly valid only for small $\beta$, 
the expression is matched to the available fixed-order computations. 
In \cite{Beneke:2011mq} the final results were based on 
the matching equation 
\begin{equation}
\label{eq:match2}
  \hat\sigma^{\text{NNLL}}_{pp'\text{matched}, 2}(\hat s)
  =\left[\hat\sigma^{\text{NNLL}}_{pp'}(\hat s)-
    \hat\sigma^{\text{NNLL}(2)}_{pp'}(\hat s)\right]
+ \hat\sigma^{\text{NLO}}_{pp'}(\hat s)
+ \hat\sigma^{\text{NNLO}}_{\text{app},pp'}(\hat s) \,,
\end{equation}
where $\hat\sigma^{\text{NNLL}(2)}_{pp'}(\hat s)$ is 
(\ref{eq:fact}) expanded to $\mathcal{O}(\alpha_s^2)$ relative 
to the Born cross section, and the last two terms represent the 
fixed-order NLO cross section plus the threshold approximation 
of the NNLO terms~\cite{Beneke:2009ye}. That is, the partonic 
cross section consisted of all known fixed-order terms with 
the $\mathcal{O}(\alpha_s^n)$ terms with $n>2$ from the 
resummation formula (\ref{eq:fact}) added.\footnote{This result was 
denoted NNLL$_2$ in \cite{Beneke:2011mq}. A second matching option, 
denoted  NNLL$_1$, was also considered. Since this implementation 
is no longer useful, once the exact NNLO result is known, we 
do not discuss it in the present paper.}

This can now be improved. The full NNLO partonic cross section, 
not restricted to small $\beta$, has recently been 
computed~\cite{Baernreuther:2012ws} for the $q\bar q$ 
initial state, except for the partonic processes 
$q\bar q\to t\bar t q\bar q$ with same-flavour (anti-)quarks 
in the initial and final state, which are presumed to be small.
The matching equation (\ref{eq:match2}) is straightforwardly 
modified to include this important result by substituting  
\begin{equation}
\label{eq:matchimpr}
\hat\sigma^{\text{NLO}}_{pp'}(\hat s)
+ \hat\sigma^{\text{NNLO}}_{\text{app},pp'}(\hat s) 
\;\to\; \hat\sigma^{\text{NNLO}}_{pp'}(\hat s)
\end{equation}
for $pp'=q\bar q$, where $\hat\sigma^{\text{NNLO}}_{pp'}(\hat s)$ 
is simply the full fixed-order NNLO cross section as given 
in~\cite{Baernreuther:2012ws}, now including 
all lower orders, and the full renormalization and factorization 
scale dependence~\cite{Langenfeld:2009wd}.\footnote{Note that 
in the expressions for $f_{q\bar q}^{21}$ and  $f_{q\bar q}^{22}$ 
in Eqs.~(A1) and (A2) of~\cite{Langenfeld:2009wd},
the functions $f_{q\bar q}^{(0)}$ multiplying the $n_f^{0,1}$ 
terms must be replaced by their threshold limit.} In the 
following this implementation will be used and denoted NNLL. 
Note, however, that we keep the labels NNLO$_{\text{app}}$ and
NNLL$_2$ for the LHC even though the full $q\bar{q}$ channel is
included as well, since at the LHC the $gg$ channel dominates and the effect
of the replacement in (\ref{eq:matchimpr}) is very small.

The theoretical uncertainties will be computed as detailed in 
\cite{Beneke:2011mq} with two modifications: (1) In the $q\bar q$ 
channel the theoretical uncertainty associated with the unknown 
NNLO result parameterized by the estimate of a constant term 
in the threshold expansion has become irrelevant. This uncertainty 
is now included only for the $gg$ channel. (2) Since we now 
include different PDF sets, we present the PDF+$\alpha_s$ error 
at the 68\% rather than 90\% confidence level (CL).

%%%%%%%%%%%%%%%%%%%%%%%%%%%%%%%%%%%%%%%%%%%%%%%%%%%%%%%%%%%%%%%%%%%%%%%%%%
\begin{table}[t]
\newcommand{\m}{\hphantom{$-$}}
\newcommand{\cc}[1]{\multicolumn{1}{c}{#1}}
\renewcommand{\tabcolsep}{0.8pc} % enlarge column spacing
\renewcommand{\arraystretch}{1.0} % enlarge line spacing
\begin{center}
\begin{tabular}{@{}llllll}
\hline  \vspace{-4mm} \\  
       \phantom{0} ${\text{NLO}}$
     & \phantom{0} ${\text{NNLO}_{\text{app}}}$
     & \phantom{0} ${\text{NNLL$_{2}$}}$  \\
\hline   \\[-1mm]
       $\phantom{0}{6.68^{\,+0.36+0.23}_{\,-0.75-0.22}}$
     & $\phantom{0}{7.06^{\,+0.24+0.10+0.29}_{\,-0.33-0.10-0.24}}$
     & $\phantom{0}{7.22^{\,+0.29+0.10+0.30}_{\,-0.46-0.10-0.25}}$
 \vspace{2mm} \\
\hline   \\[-1mm]
       
     & $\phantom{0}{7.00^{\,+0.21+0.02+0.29}_{\,-0.31-0.02-0.25}}$
     & $\phantom{0}{7.15^{\,+0.21+0.02+0.30}_{\,-0.20-0.02-0.25}}$
 \vspace{2mm} \\
\hline \\[-4mm]
       \phantom{ab} 
     & \phantom{0}${\text{NNLO}}$
     & \phantom{0}${\text{NNLL}}$  \\
\hline
\end{tabular}\\
\caption{\sf The total top-pair cross section (in pb) 
for $m_{t}=173.3$~GeV in $p\bar p$ collisions at the Tevatron 
($\sqrt{s}=1.96\,$TeV) with MSTW2008NNLO (NLO for NLO) input. 
The first set of errors refers to scale variation (scale 
variation+resummation ambiguities for NNLL$_2$ and NNLL),
the last to the 68\% CL PDF+$\alpha_s$ error. The second set of errors for 
NNLO$_{(\text{app})}$/NNLL$_{(2)}$ arises from
variations of the unknown NNLO constant term.}
\label{tab:tevatron173.3}
\end{center}
\end{table} 

%%%%%%%%%%%%%%%%%%%%%%%%%%%%%%%%%%%%%%%%
\subsection{Tevatron results}
Since top pairs in $p\bar p$ collisions at the Tevatron are produced
predominantly through the $q\bar q$ initial state, we are now in the
position to present a (nearly complete) NNLL result matched to the
(nearly complete) NNLO fixed-order calculation. We update our results
from \cite{Beneke:2011mq} using the same top-quark pole mass
$m_t=173.3\,$GeV and MSTW2008NNLO~\cite{Martin:2009iq} PDF input with
$\alpha_s(M_Z) = 0.1171^{+0.0014}_{-0.0014}$~\cite{Martin:2009bu}.
We compare the various approximations in
Table~\ref{tab:tevatron173.3}. The NLO,
NNLO$_{\text{app}}$ and NNLL$_{2}$ entries are the same as in
\cite{Beneke:2011mq} except for the PDF+$\alpha_s$ error which is now
given at 68\% rather than 90\%~CL. With the new NNLO result for the
$q\bar q$ channel included (second line in the Table), the cross
section is reduced by about $1\%$.\footnote{Note that the difference
  between NNLO and NNLO$_{\text{app}}$ on the one hand, and NNLL and
  NNLL$_{2}$ on the other is not exactly the same, since we recompute
  the value of $\beta_{\rm cut}$ in the resummation method 2, see
  \cite{Beneke:2011mq}, which changes from 0.35 for NNLL$_{2}$ to 0.38
  for NNLL.}  The theoretical uncertainty of the resummed cross
section from scale and resummation ambiguities and the unknown
constant is reduced from $\pm 5.4\%$ to $\pm 2.9\%$.
The reduction of the  error from NNLO$_{\text{app}}$ to NNLO is
smaller which we attribute to the fact that a non-negligible part of
the scale uncertainty originates from the gluon-gluon initial state
(see Table 5 in \cite{Beneke:2011mq}) so a further reduction is
expected from the full NNLO calculation for all initial states.
 The smallness of
the correction shows {\em a posteriori} that the threshold
approximation works rather well (though, partly, by a fortuitous
cancellation in the large $\beta$ region,
see~\cite{Baernreuther:2012ws}). Our best result for the Tevatron $t\bar t$
production cross section, assuming $m_t=173.3\,$GeV and 
$\alpha_s(M_Z)=0.1171$, is therefore
\begin{equation}
\label{bestTevatron}
\sigma_{t\bar t} = 7.15^{\,+0.21}_{\,-0.20}\,\mbox{(theory)}
{}^{\,+0.30}_{\,-0.25} \,\mbox{(PDF+$\alpha_s$)} \,\mbox{pb}
\qquad (\mbox{MSTW2008NNLO}).
\end{equation}
The cross section increases to $7.26\,$pb, see Table~\ref{tab:xs_all} below, 
when the ``world average'' value $\alpha_s(M_Z)=0.1180$ is used.
We estimate the theory uncertainty by adding the errors from scale
variation, resummation ambiguities and the estimate of the constant 
NNLO term in the gluon channel in quadrature, where the latter has 
negligible effect at the Tevatron. 
Table~\ref{tab:mt-tev} provides the corresponding 
results in the range $m_t=165$--$180$~GeV. 

%%%%%%%%%%%%%%%%%%%%%%%%%%%%%%%%%%%%%%%%%%%%%%%%%%%%%%%%%%%%%%%%%%%%%%%%%%
\begin{table}[!t]
\newcommand{\m}{\hphantom{$-$}}
\newcommand{\cc}[1]{\multicolumn{1}{c}{#1}}
\renewcommand{\tabcolsep}{0.8pc} % enlarge column spacing
\renewcommand{\arraystretch}{1.0} % enlarge line spacing
\caption{ \sf Total cross sections in pb at the Tevatron for 
$m_{t}=165$--$180$~GeV. See caption of Table~\ref{tab:tevatron173.3} 
for the definition of the various errors.}
\label{tab:mt-tev}
\begin{center}
\begin{tabular}{llll}
\hline\\[-4.3mm]  
$m_t$~[GeV]  &${\text{NLO}}$&${\text{NNLO}}$&${\text{NNLL}}$
\\\hline\\[-3mm] 
      165
     & $\phantom{0}{{8.70^{\,+0.48+0.30}_{\,-0.98-0.30}}}$
     & $\phantom{0}{{9.10^{\,+0.27+0.04+0.38}_{\,-0.40-0.04-0.32}}}$
     & $\phantom{0}{{9.29^{\,+0.29+0.04+0.40}_{\,-0.26-0.04-0.33}}}$
      \vspace{2mm} \\
      166
     & $\phantom{0}{{8.42^{\,+0.46+0.29}_{\,-0.95-0.29}}}$
     & $\phantom{0}{{8.82^{\,+0.26+0.03+0.37}_{\,-0.38-0.03-0.31}}}$
     & $\phantom{0}{{9.00^{\,+0.28+0.03+0.38}_{\,-0.25-0.03-0.32}}}$
      \vspace{2mm} \\
      167
     & $\phantom{0}{{8.16^{\,+0.45+0.28}_{\,-0.92-0.28}}}$
     & $\phantom{0}{{8.54^{\,+0.25+0.03+0.36}_{\,-0.37-0.03-0.30}}}$
     & $\phantom{0}{{8.71^{\,+0.27+0.03+0.37}_{\,-0.24-0.03-0.31}}}$
      \vspace{2mm} \\
      168
     & $\phantom{0}{{7.90^{\,+0.43+0.27}_{\,-0.89-0.27}}}$
     & $\phantom{0}{{8.27^{\,+0.25+0.03+0.35}_{\,-0.36-0.03-0.29}}}$
     & $\phantom{0}{{8.44^{\,+0.26+0.03+0.36}_{\,-0.24-0.03-0.30}}}$
      \vspace{2mm} \\
      169
     & $\phantom{0}{{7.65^{\,+0.42+0.26}_{\,-0.86-0.26}}}$
     & $\phantom{0}{{8.01^{\,+0.24+0.03+0.33}_{\,-0.35-0.03-0.28}}}$
     & $\phantom{0}{{8.18^{\,+0.25+0.03+0.34}_{\,-0.23-0.03-0.29}}}$
      \vspace{2mm} \\
      170
     & $\phantom{0}{{7.41^{\,+0.40+0.26}_{\,-0.84-0.25}}}$
     & $\phantom{0}{{7.76^{\,+0.23+0.03+0.32}_{\,-0.34-0.03-0.27}}}$
     & $\phantom{0}{{7.92^{\,+0.24+0.03+0.33}_{\,-0.22-0.03-0.28}}}$
      \vspace{2mm} \\
      171
     & $\phantom{0}{{7.18^{\,+0.39+0.25}_{\,-0.81-0.24}}}$
     & $\phantom{0}{{7.52^{\,+0.22+0.03+0.31}_{\,-0.33-0.03-0.26}}}$
     & $\phantom{0}{{7.68^{\,+0.23+0.03+0.32}_{\,-0.22-0.03-0.27}}}$
      \vspace{2mm} \\
      172
     & $\phantom{0}{{6.96^{\,+0.38+0.24}_{\,-0.78-0.23}}}$
     & $\phantom{0}{{7.29^{\,+0.22+0.03+0.30}_{\,-0.32-0.03-0.25}}}$
     & $\phantom{0}{{7.44^{\,+0.22+0.03+0.31}_{\,-0.21-0.03-0.26}}}$
      \vspace{2mm} \\
      173
     & $\phantom{0}{{6.74^{\,+0.37+0.23}_{\,-0.76-0.23}}}$
     & $\phantom{0}{{7.07^{\,+0.21+0.02+0.29}_{\,-0.31-0.02-0.24}}}$
     & $\phantom{0}{{7.21^{\,+0.21+0.02+0.30}_{\,-0.20-0.02-0.25}}}$
      \vspace{2mm} \\
      174
     & $\phantom{0}{{6.54^{\,+0.35+0.23}_{\,-0.74-0.22}}}$
     & $\phantom{0}{{6.85^{\,+0.20+0.02+0.28}_{\,-0.30-0.02-0.23}}}$
     & $\phantom{0}{{6.99^{\,+0.21+0.02+0.29}_{\,-0.20-0.02-0.24}}}$
      \vspace{2mm} \\
      175
     & $\phantom{0}{{6.34^{\,+0.34+0.22}_{\,-0.71-0.21}}}$
     & $\phantom{0}{{6.64^{\,+0.20+0.02+0.27}_{\,-0.29-0.02-0.22}}}$
     & $\phantom{0}{{6.78^{\,+0.20+0.02+0.28}_{\,-0.19-0.02-0.23}}}$
      \vspace{2mm} \\
      176
     & $\phantom{0}{{6.14^{\,+0.33+0.21}_{\,-0.69-0.20}}}$
     & $\phantom{0}{{6.44^{\,+0.19+0.02+0.26}_{\,-0.29-0.02-0.22}}}$
     & $\phantom{0}{{6.57^{\,+0.19+0.02+0.27}_{\,-0.19-0.02-0.22}}}$
      \vspace{2mm} \\
      177
     & $\phantom{0}{{5.96^{\,+0.32+0.21}_{\,-0.67-0.20}}}$
     & $\phantom{0}{{6.24^{\,+0.19+0.02+0.25}_{\,-0.28-0.02-0.21}}}$
     & $\phantom{0}{{6.38^{\,+0.19+0.02+0.26}_{\,-0.18-0.02-0.22}}}$
      \vspace{2mm} \\
      178
     & $\phantom{0}{{5.78^{\,+0.31+0.20}_{\,-0.65-0.19}}}$
     & $\phantom{0}{{6.06^{\,+0.18+0.02+0.25}_{\,-0.27-0.02-0.20}}}$
     & $\phantom{0}{{6.18^{\,+0.18+0.02+0.25}_{\,-0.18-0.02-0.21}}}$
      \vspace{2mm} \\
      179
     & $\phantom{0}{{5.60^{\,+0.30+0.20}_{\,-0.63-0.18}}}$
     & $\phantom{0}{{5.87^{\,+0.17+0.02+0.24}_{\,-0.26-0.02-0.20}}}$
     & $\phantom{0}{{6.00^{\,+0.17+0.02+0.25}_{\,-0.17-0.02-0.20}}}$
      \vspace{2mm} \\
      180
     & $\phantom{0}{{5.43^{\,+0.29+0.19}_{\,-0.61-0.18}}}$
     & $\phantom{0}{{5.70^{\,+0.17+0.02+0.23}_{\,-0.25-0.02-0.19}}}$
     & $\phantom{0}{{5.82^{\,+0.17+0.02+0.24}_{\,-0.17-0.02-0.20}}}$ \\[1mm]
\hline
\end{tabular}\\
\end{center}
\end{table} 
%%%%%%%%%%%%%%%%%%%%%%%%%%%%%%%%%%%%%%%%%%%%%%%%

It is instructive to compare the effect of adding the full NNLO $q\bar q$ 
partonic cross section to the previous NNLL resummations in the 
momentum-space or SCET-based resummation formalism employed here and in 
\cite{Beneke:2011mq}, and to the Mellin-space 
formalism~\cite{Cacciari:2011hy,Czakon:2011xx}. This is particularly 
so as the Mellin-space result as well as the result 
from \cite{Ahrens:2011px} have been somewhat lower than those 
given in \cite{Beneke:2011mq}, see the comparison \cite{Beneke:2011ys}. 
For the central values, we find
\[
\begin{array}{lllll}
7.22\,\mbox{pb} \;\mbox{($\!\!$\cite{Beneke:2011mq})} & \;\to\;& 
7.15\,\mbox{pb} \;\mbox{(Eq. (\ref{bestTevatron}))}
&\quad& (\mbox{momentum-space}),
\\[0.2cm]
6.72\,\mbox{pb} \;\mbox{($\!\!$\cite{Cacciari:2011hy})} &\;\to\;& 
7.07\,\mbox{pb} \;\mbox{($\!\!$\cite{Baernreuther:2012ws})} 
&\quad& \mbox{(Mellin-space)},
\end{array} 
\]
for identical $m_t$, $\alpha_s$ and PDF input. We observe that the two
approaches are now in full agreement, and that the momentum-space
result has turned out to be substantially less sensitive to the
inclusion of the full NNLO $q\bar q$ channel than the Mellin-space
one\footnote{This is partly due to the change in the matching
  procedure in the Mellin-space results (from NLO+NNLL in
  \cite{Cacciari:2011hy} to NNLO+NNLL in \cite{Baernreuther:2012ws})
  and the different treatment of the NNLO constant term in
  \cite{Beneke:2011mq} and \cite{Cacciari:2011hy}.}.
Furthermore, the comparison to the full NNLO result shows 
that approximate NNLO expressions obtained from integrating approximate 
invariant mass or one-particle inclusive 
distributions~\cite{Ahrens:2011px} do not lead to a better 
approximation than the threshold expansion. 

It is interesting to study the impact of the new, almost complete,
NNLO+NNLL predictions on the extraction of the pole mass of the top
quark from the measured total cross section at the Tevatron. The D0
collaboration~\cite{Abazov:2011pt} extracted the pole and
$\overline{\text{MS}}$ mass using several higher-order predictions of
the cross section. For instance, the pole mass $m_t=167.5^{+
  5.2}_{-4.7}$~GeV was obtained using the approximate NNLO calculation
from~\cite{Langenfeld:2009wd} and MSTW2008 PDFs. We now study the
impact of our updated theory predictions on the mass measurement. 
 To this end we fit the NNLL cross section in
Table~\ref{tab:mt-tev} by a function of the form
\begin{equation} \label{eq:theoryxs}
\sigma_{t \bar{t}}^{\text{th}}(m_t) = 
\left(\frac{172.5}{m_t}\right)^4 
\left(c_0+c_1 (m_t-172.5)+c_2 (m_t-172.5)^2+c_3 (m_t-172.5)^3 \right) \, 
\text{pb}\, ,
\end{equation}
where all the masses are given in GeV.  For the coefficients in the
fit~\eqref{eq:theoryxs} we find $c_0=7.325\pm 0.219\pm 0.304$,
$c_1=-(5.8651\pm 0.3163\pm 0.3382)\times 10^{-2}$, $c_2=(2.4884\pm
0.7131 \pm 0.9027)\times 10^{-4}$, $c_3=(3.7436\pm 4.1254\pm
1.9924)\times 10^{-6}$ where the first error denotes the total theory
error and the second the PDF$+\alpha_s$ error.  The maximum of the
positive and negative errors has been used in the fit.  An analogous
fit for the dependence of the experimental cross section on the
reference mass is given by the D0 collaboration
in~\cite{Abazov:2011cq} for the combination of the dilepton and
lepton+jet results that corresponds to $\sigma_{t\bar t}=
7.56^{+0.63}_{-0.56}$ pb for a reference mass of $m_t=172.5$~GeV.  As
in~\cite{Beneke:2011mq} the most probable value of the pole mass is
obtained by the maximization of a joined likelihood function obtained
from a convolution of two normalized Gaussians centered at the
theoretical prediction and the experimental mean value, respectively,
and with a width given by the corresponding errors. This method is
similar to the one used in~\cite{Abazov:2011pt}, however, we do not
introduce separate likelihood functions for scale and PDF uncertainty,
but add theory and PDF errors linearly.  Using our best NNLL
prediction, we obtain\footnote{The uncertainty is not improved
  compared to~\cite{Abazov:2011pt} that used selection criteria
  resulting in a weaker $m_t$-dependence of the experimental cross
  section than in~\cite{Abazov:2011cq}, but a larger cross section
  measurement of $\sigma_{t\bar t}= 8.13^{+1.02}_{-0.90}$ pb for a
  reference mass of $m_t=172.5$~GeV.  Naively rescaling the
  mass-dependence of the experimental cross section
  in~\cite{Abazov:2011pt} to the combined central value and error
  of~\cite{Abazov:2011cq} we obtain
  $m_t=171.6^{+4.3}_{-4.3}$~\text{GeV} which might indicate the
  precision that could be achieved with the Tevatron data.}
\begin{equation}
\label{eq:mt-TeV}
  m_t=171.4^{+5.4}_{-5.7}\, \text{GeV}.  
\end{equation}
This result is in good agreement with the combined result
$m_t=173.2\pm 0.8$~GeV from the direct mass determination at the
Tevatron~\cite{Lancaster:2011wr} as well as our previous
result~\cite{Beneke:2011mq} $m_t = 169.8^{+4.9}_{-4.7}$ GeV using an
ATLAS cross-section measurement~\cite{ATLAS-CONF-2011-121} and the value $m_t =
170.3^{+7.3}_{-6.7}$~GeV obtained from a CMS
cross-section measurement~\cite{Aldaya:2012tv} using the calculation
of~\cite{Langenfeld:2009wd}. In~\eqref{eq:mt-TeV} the mass used in the
kinematic reconstruction of the top-events is identified with the
pole mass. Allowing for a difference of $\pm 1$~GeV in the two mass
definitions leads to an additional uncertainty of $0.5$~GeV.  To study
the effect of the exact NNLO result of~\cite{Baernreuther:2012ws} and
of NNLL resummation, we have repeated the same procedure for the
approximate NNLO results presented in~\cite{Beneke:2011mq} and the
NNLO results in Table~\ref{tab:mt-tev}.  We obtain $
m_t=171.0^{+5.8}_{-6.3}$ for the approximate NNLO calculation and $
m_t=170.5^{+5.7}_{-6.4}$ for the full NNLO result,
respectively. Consistent with Table~\ref{tab:tevatron173.3}, the effect
of the result of~\cite{Baernreuther:2012ws} on the NNLO calculation is
moderate, while a stronger reduction of the error is observed for the
NNLL prediction.

\subsection{\boldmath LHC results}
\label{sec:8tev}

%%%%%%%%%%%%%%%%%%%%%%%%%%%%%%%%%%%%%%%%%%%%%%%%%%%%%%%%%%%%%%%%%%%%%%%%%%
\begin{table}[!tb]
\caption{ \sf Total cross sections in pb at the LHC ($\sqrt{s}=7$ TeV)
  for $m_{t}=165$--$180$~GeV. See caption of
  Table~\ref{tab:tevatron173.3} for the definition of the various
  errors.}
\label{tab:mt-lhc7}
\newcommand{\m}{\hphantom{$-$}}
\newcommand{\cc}[1]{\multicolumn{1}{c}{#1}}
\renewcommand{\tabcolsep}{0.8pc} % enlarge column spacing
\renewcommand{\arraystretch}{1.0} % enlarge line spacing
\begin{center}
\begin{tabular}{llll}
\hline  \\[-4.3mm] 
$m_t$~[GeV]  &${\text{NLO}}$&${\text{NNLO}_{\text{app}}}$&${\text{NNLL$_{2}$}}$
\\\hline\\[-3mm] 
      165
     & $\phantom{0}{{203.9^{\,+25.5+8.6}_{\,-27.4-7.8}}}$
     & $\phantom{0}{{207.4^{\,+13.6+5.7+9.2}_{\,-14.0-5.7-8.5}}}$
     & $\phantom{0}{{209.4^{\,+6.6+5.7+9.3}_{\,-6.9-5.7-8.6}}}$
      \vspace{2mm} \\
      166
     & $\phantom{0}{{197.7^{\,+24.7+8.4}_{\,-26.5-7.6}}}$
     & $\phantom{0}{{201.1^{\,+13.1+5.5+8.9}_{\,-13.6-5.5-8.2}}}$
     & $\phantom{0}{{202.9^{\,+6.4+5.5+9.0}_{\,-6.7-5.5-8.3}}}$
      \vspace{2mm} \\
      167
     & $\phantom{0}{{191.6^{\,+23.9+8.1}_{\,-25.7-7.4}}}$
     & $\phantom{0}{{195.0^{\,+12.7+5.3+8.6}_{\,-13.1-5.3-8.0}}}$
     & $\phantom{0}{{196.8^{\,+6.3+5.3+8.7}_{\,-6.5-5.3-8.1}}}$
      \vspace{2mm} \\
      168
     & $\phantom{0}{{185.8^{\,+23.1+7.9}_{\,-24.9-7.2}}}$
     & $\phantom{0}{{189.0^{\,+12.2+5.1+8.4}_{\,-12.7-5.1-7.8}}}$
     & $\phantom{0}{{190.8^{\,+6.1+5.1+8.5}_{\,-6.3-5.1-7.9}}}$
      \vspace{2mm} \\
      169
     & $\phantom{0}{{180.2^{\,+22.4+7.7}_{\,-24.2-7.0}}}$
     & $\phantom{0}{{183.3^{\,+11.8+5.0+8.1}_{\,-12.3-5.0-7.5}}}$
     & $\phantom{0}{{185.0^{\,+5.9+5.0+8.3}_{\,-6.1-5.0-7.6}}}$
      \vspace{2mm} \\
      170
     & $\phantom{0}{{174.7^{\,+21.7+7.5}_{\,-23.4-6.8}}}$
     & $\phantom{0}{{177.8^{\,+11.4+4.8+7.9}_{\,-11.9-4.8-7.3}}}$
     & $\phantom{0}{{179.5^{\,+5.7+4.8+8.0}_{\,-5.9-4.8-7.4}}}$
      \vspace{2mm} \\
      171
     & $\phantom{0}{{169.5^{\,+21.0+7.2}_{\,-22.7-6.6}}}$
     & $\phantom{0}{{172.5^{\,+11.0+4.6+7.7}_{\,-11.5-4.6-7.1}}}$
     & $\phantom{0}{{174.1^{\,+5.5+4.6+7.8}_{\,-5.8-4.6-7.2}}}$
      \vspace{2mm} \\
      172
     & $\phantom{0}{{164.4^{\,+20.3+7.0}_{\,-22.0-6.4}}}$
     & $\phantom{0}{{167.4^{\,+10.7+4.5+7.5}_{\,-11.1-4.5-6.9}}}$
     & $\phantom{0}{{168.9^{\,+5.4+4.5+7.6}_{\,-5.6-4.5-7.0}}}$
      \vspace{2mm} \\
      173
     & $\phantom{0}{{159.6^{\,+19.7+6.8}_{\,-21.4-6.2}}}$
     & $\phantom{0}{{162.4^{\,+10.3+4.3+7.3}_{\,-10.8-4.3-6.7}}}$
     & $\phantom{0}{{163.9^{\,+5.2+4.3+7.4}_{\,-5.4-4.3-6.8}}}$
      \vspace{2mm} \\
      174
     & $\phantom{0}{{154.8^{\,+19.1+6.6}_{\,-20.7-6.1}}}$
     & $\phantom{0}{{157.6^{\,+10.0+4.2+7.1}_{\,-10.5-4.2-6.5}}}$
     & $\phantom{0}{{159.1^{\,+5.1+4.2+7.3}_{\,-5.3-4.2-6.6}}}$
      \vspace{2mm} \\
      175
     & $\phantom{0}{{150.3^{\,+18.5+6.5}_{\,-20.1-5.9}}}$
     & $\phantom{0}{{153.0^{\,+9.7+4.1+6.9}_{\,-10.1-4.1-6.4}}}$
     & $\phantom{0}{{154.4^{\,+4.9+4.1+7.0}_{\,-5.1-4.1-6.4}}}$
      \vspace{2mm} \\
      176
     & $\phantom{0}{{145.9^{\,+17.9+6.3}_{\,-19.5-5.7}}}$
     & $\phantom{0}{{148.5^{\,+9.4+3.9+6.7}_{\,-9.8-3.9-6.2}}}$
     & $\phantom{0}{{149.9^{\,+5.3+3.9+6.8}_{\,-5.0-3.9-6.3}}}$
      \vspace{2mm} \\
      177
     & $\phantom{0}{{141.6^{\,+17.4+6.1}_{\,-19.0-5.6}}}$
     & $\phantom{0}{{144.2^{\,+9.1+3.8+6.5}_{\,-9.5-3.8-6.0}}}$
     & $\phantom{0}{{145.5^{\,+4.6+3.8+6.6}_{\,-4.9-3.8-6.1}}}$
      \vspace{2mm} \\
      178
     & $\phantom{0}{{137.5^{\,+16.9+5.9}_{\,-18.4-5.4}}}$
     & $\phantom{0}{{140.0^{\,+8.8+3.7+6.3}_{\,-9.2-3.7-5.9}}}$
     & $\phantom{0}{{141.3^{\,+4.5+3.7+6.4}_{\,-4.7-3.7-5.9}}}$
      \vspace{2mm} \\
      179
     & $\phantom{0}{{133.5^{\,+16.3+5.8}_{\,-17.9-5.3}}}$
     & $\phantom{0}{{136.0^{\,+8.5+3.6+6.1}_{\,-8.9-3.6-5.7}}}$
     & $\phantom{0}{{137.3^{\,+4.4+3.6+6.2}_{\,-4.6-3.6-5.8}}}$
      \vspace{2mm} \\
      180
     & $\phantom{0}{{129.7^{\,+15.9+5.6}_{\,-17.3-5.1}}}$
     & $\phantom{0}{{132.1^{\,+8.2+3.5+6.0}_{\,-8.7-3.5-5.5}}}$
     & $\phantom{0}{{133.3^{\,+4.3+3.5+6.1}_{\,-4.5-3.5-5.6}}}$ \\[1mm]
\hline
\end{tabular}\\
\end{center}
\end{table} 
%%%%%%%%%%%%%%%%%%%%%%%%%%%%%%%%%%%%%%%%%%%%%%%%%%%%%%%%%%%%%%%%%%%%%%%%%%

In this section we provide predictions for the top-antitop total cross
section, computed with \prog, for the LHC with the current
centre-of-mass energy of $8\,$TeV, extending the results for $7$ and
$14$~TeV given in~\cite{Beneke:2011mq} (the result for $m_t=173.3$~GeV
with the setup of~\cite{Beneke:2011mq} was already presented
in~\cite{Beneke:2012eb}). We also update the results for $7\,$TeV
centre-of-mass energy by including the full NNLO result for the $q\bar
q$-initial state, although the effect on the central value is
negligible since $t \bar{t}$ production at the LHC is dominated by the
gluon-fusion channel where the full NNLO corrections are not known
yet. Since the LHC centre-of-mass
energy after the 2013--2014 shutdown has not been fixed, we do not give
results for higher centre-of-mass energies.
Results for other centre-of-mass energies can be easily obtained with
{\prog} since $\sqrt{s}$ can be freely chosen in the configuration file,
see the Appendix.  Our best result for the $t\bar t$
production cross section at the LHC, using $m_t=173.3\,$GeV, the
MSTW2008 PDF sets~\cite{Martin:2009iq} and $\alpha_s(M_Z)=0.1171$, 
is given by
\begin{align}
\label{bestL7}
\sigma_{t\bar t}\, (\text{LHC}, 7 \,\text{TeV})&=
162.4^{\,+6.7}_{\,-6.9}\,\mbox{(theory)}
{}^{\,+7.3}_{\,-6.8} \,\mbox{(PDF+$\alpha_s$)} \,\mbox{pb}
\qquad (\mbox{MSTW2008NNLO})\\[0.3cm]
\sigma_{t\bar t} (\text{LHC}, 8\, \text{TeV})&=
231.8^{\,+9.6}_{\,-9.9}\,\mbox{(theory)} {}^{\,+9.8}_{\,-9.1}
\,\mbox{(PDF+$\alpha_s$)} \,\mbox{pb}
\qquad (\mbox{MSTW2008NNLO})
\label{bestL8}
\end{align} 
The NLO, NNLO and NNLL $t \bar{t}$ cross sections at the LHC with
$\sqrt{s}=7,\, 8\,$TeV for the mass range $m_t=165$--$180\,$GeV are shown
in Table \ref{tab:mt-lhc7} and \ref{tab:mt-lhc8}, respectively.  The
effect of corrections beyond the next-to-leading order at the LHC with
$\sqrt{s}=8\,$TeV is similar to that observed for $\sqrt{s}=7\,$TeV,
increasing the cross section by $2.4\%$ compared to the NLO
result, with fixed-order NNLO contributions accounting for $1.6\%$.
It can be seen from Tables \ref{tab:mt-lhc7} and
\ref{tab:mt-lhc8} that the uncertainty from the unknown fixed-order
NNLO corrections for the $gg$ (and $qg$) initial states, as estimated
by the variation of the constant term in the threshold expansion, are
of the same order as the remaining theory errors of the NNLL results
at the LHC. This source of uncertainty will be removed once the full
NNLO calculation has been performed also for these
channels\footnote{For recent work constraining the NNLO cross
  section using information on the $\beta\to 1$ or large
  invariant-mass limits, see~\cite{Moch:2012mk,Ferroglia:2012ku}.}.

%%%%%%%%%%%%%%%%%%%%%%%%%%%%%%%%%%%%%%%%%%%%%%%%%%%%%%%%%%%%%%%%%%%%%%%%%%
\begin{table}[!tb]
\caption{ \sf Total cross sections in pb at the LHC ($\sqrt{s}=8$ TeV)
  for $m_{t}=165$--$180$~GeV. See caption of
  Table~\ref{tab:tevatron173.3} for the definition of the various
  errors.}
\label{tab:mt-lhc8}
\newcommand{\m}{\hphantom{$-$}}
\newcommand{\cc}[1]{\multicolumn{1}{c}{#1}}
\renewcommand{\tabcolsep}{0.8pc} % enlarge column spacing
\renewcommand{\arraystretch}{1.0} % enlarge line spacing
\begin{center}
\begin{tabular}{llll}
\hline  \\[-4.3mm] 
$m_t$~[GeV]  &${\text{NLO}}$&${\text{NNLO}_{\text{app}}}$&${\text{NNLL$_{2}$}}$
\\\hline\\[-3mm] 
      165
     & $\phantom{0}{{290.1^{\,+36.1+11.7}_{\,-38.1-10.4}}}$
     & $\phantom{0}{{294.5^{\,+20.0+8.3+12.3}_{\,-20.1-8.3-11.4}}}$
     & $\phantom{0}{297.0_ {-9.6}^{+9.1}{}_{-8.3}^{+8.3}{}_{-11.5}^{+12.4}}$
      \vspace{2mm} \\
      166
     & $\phantom{0}{{281.4^{\,+35.0+11.4}_{\,-37.0-10.1}}}$
     & $\phantom{0}{{285.7^{\,+19.4+8.0+12.0}_{\,-19.5-8.0-11.0}}}$
     & $\phantom{0}{{288.1_{-9.3}^{+8.9}{}_{-8.0}^{+8.0}{}_{-11.2}^{+12.1}}}$
      \vspace{2mm} \\
      167
     & $\phantom{0}{{273.0^{\,+33.9+11.1}_{\,-35.9-9.8}}}$
     & $\phantom{0}{{277.2^{\,+18.7+7.8+11.6}_{\,-18.8-7.8-10.7}}}$
     & $\phantom{0}{{279.5_{-9.1}^{+8.6}{}_{-7.8}^{+7.8}{}_{-10.9}^{+11.8}}}$
      \vspace{2mm} \\
      168
     & $\phantom{0}{{264.9^{\,+32.8+10.8}_{\,-34.8-9.6}}}$
     & $\phantom{0}{{269.0^{\,+18.1+7.5+11.2}_{\,-18.3-7.5-10.4}}}$
     & $\phantom{0}{{271.2_{-8.8}^{+8.4}{}_{-7.5}^{+7.5}{}_{-10.6}^{+11.4}}}$
      \vspace{2mm} \\
      169
     & $\phantom{0}{{257.0^{\,+31.8+10.4}_{\,-33.7-9.3}}}$
     & $\phantom{0}{{261.0^{\,+17.5+7.3+11.0}_{\,-17.7-7.3-10.2}}}$
     & $\phantom{0}{{263.2_ {-8.6}^{+8.1}{}_{-7.3}^{+7.3}{}_{-10.3}^{+11.1}}}$
      \vspace{2mm} \\
      170
     & $\phantom{0}{{249.5^{\,+30.8+10.2}_{\,-32.7-9.0}}}$
     & $\phantom{0}{{253.3^{\,+17.0+7.0+10.7}_{\,-17.1-7.0-9.9}}}$
     & $\phantom{0}{{255.5_ {-8.3}^{+7.9}{}_{-7.0}^{+7.0}{}_{-10.0}^{+10.8}}}$
      \vspace{2mm} \\
      171
     & $\phantom{0}{{242.1^{\,+29.9+9.9}_{\,-31.8-8.8}}}$
     & $\phantom{0}{{245.9^{\,+16.4+6.8+10.4}_{\,-16.6-6.8-9.6}}}$
     & $\phantom{0}{{248.0_{-8.1}^{+7.6}{}_{-6.8}^{+6.8}{}_{-9.7}^{+10.5}}}$
      \vspace{2mm} \\
      172
     & $\phantom{0}{{235.1^{\,+29.0+9.6}_{\,-30.8-8.6}}}$
     & $\phantom{0}{{238.8^{\,+15.9+6.6+10.1}_{\,-16.1-6.6-9.3}}}$
     & $\phantom{0}{{240.8_{-7.9}^{+7.4}{}_{-6.6}^{+6.6}{}_{-9.4}^{+10.2}}}$
      \vspace{2mm} \\
      173
     & $\phantom{0}{{228.3^{\,+28.1+9.3}_{\,-29.9-8.3}}}$
     & $\phantom{0}{{231.9^{\,+15.4+6.4+9.8}_{\,-15.6-6.4-9.1}}}$
     & $\phantom{0}{{233.9_{-7.6}^{+7.2}{}_{-6.4}^{+6.4}{}_{-9.2}^{+9.9}}}$
      \vspace{2mm} \\
      174
     & $\phantom{0}{{221.7^{\,+27.3+9.1}_{\,-29.1-8.1}}}$
     & $\phantom{0}{{225.2^{\,+14.9+6.2+9.5}_{\,-15.1-6.2-8.8}}}$
     & $\phantom{0}{{227.1_{-7.4}^{+7.0}{}_{-6.2}^{+6.2}{}_{-8.9}^{+9.7}}}$
      \vspace{2mm} \\
      175
     & $\phantom{0}{{215.3^{\,+26.4+8.8}_{\,-28.2-7.9}}}$
     & $\phantom{0}{{218.7^{\,+14.4+6.0+9.3}_{\,-14.6-6.0-8.6}}}$
     & $\phantom{0}{{220.6_{-7.2}^{+6.8}{}_{-6.0}^{+6.0}{}_{-8.7}^{+9.4}}}$
      \vspace{2mm} \\
      176
     & $\phantom{0}{{209.1^{\,+25.6+8.6}_{\,-27.4-7.7}}}$
     & $\phantom{0}{{212.5^{\,+14.0+5.8+9.0}_{\,-14.2-5.8-8.4}}}$
     & $\phantom{0}{{214.3_{-7.0}^{+6.6}{}_{-5.8}^{+5.8}{}_{-8.5}^{+9.1}}}$
      \vspace{2mm} \\
      177
     & $\phantom{0}{{203.2^{\,+24.9+8.3}_{\,-26.6-7.5}}}$
     & $\phantom{0}{{206.4^{\,+13.5+5.6+8.8}_{\,-13.8-5.6-8.1}}}$
     & $\phantom{0}{{208.2_{-6.8}^{+6.4}{}_{-5.6}^{+5.6}{}_{-8.2}^{+8.9}}}$
      \vspace{2mm} \\
      178
     & $\phantom{0}{{197.4^{\,+24.1+8.1}_{\,-25.9-7.3}}}$
     & $\phantom{0}{{200.6^{\,+13.1+5.5+8.6}_{\,-13.3-5.5-7.9}}}$
     & $\phantom{0}{{202.3_{-6.7}^{+6.2}{}_{-5.5}^{+5.5}{}_{-8.0}^{+8.7}}}$
      \vspace{2mm} \\
      179
     & $\phantom{0}{{191.8^{\,+23.4+7.9}_{\,-25.1-7.1}}}$
     & $\phantom{0}{{194.9^{\,+12.7+5.3+8.3}_{\,-12.9-5.3-7.7}}}$
     & $\phantom{0}{{196.6_{-6.5}^{+6.1}{}_{-5.3}^{+5.3}{}_{-7.8}^{+8.4}}}$
      \vspace{2mm} \\
      180
     & $\phantom{0}{{186.4^{\,+22.7+7.7}_{\,-24.4-6.9}}}$
     & $\phantom{0}{{189.5^{\,+12.3+5.1+8.1}_{\,-12.6-5.1-7.5}}}$
     & $\phantom{0}{{191.1_{-6.3}^{+5.9}{}_{-5.1}^{+5.1}{}_{-7.6}^{+8.2}}}$ \\[1mm]
\hline
\end{tabular}
\end{center}
\end{table} 
%%%%%%%%%%%%%%%%%%%%%%%%%%%%%%%%%%%%%%%%%%%%%%%%%%%%%%%%%%%%%%%%%%%%%%%%%%
Comparing to our previous predictions for the LHC with $\sqrt
s=7\,$TeV~\cite{Beneke:2011mq} and $\sqrt
s=8\,$TeV~\cite{Beneke:2012eb} it is seen that in our approach the
inclusion of the exact NNLO result for the $q \bar{q}$ channel amounts
to a minimal change of the central value by $-(0.1$--$0.2)$~pb:
\[
\begin{array}{lllll}
162.6\,\mbox{pb} \;\mbox{($\!\!$\cite{Beneke:2011mq})} & \;\to\;& 
162.4\,\mbox{pb} \;\mbox{(Eq. (\ref{bestL7}))} 
&\quad& \text{($7$~TeV)}, \\[0.2cm]
231.9\,\mbox{pb} \;\mbox{($\!\!$\cite{Beneke:2012eb})}
 & \;\to\;&
 231.8\,\mbox{pb}
 \;\mbox{(Eq. (\ref{bestL8}))}
 &\quad& \text{($8$~TeV)}. 
\end{array} 
\]
The effect on the theory error is somewhat larger and we find a
reduction from ${}^{+7.4}_{-7.6}$ to ${}^{+6.7}_{-6.9}$~pb at $\sqrt
s=7$~TeV and from ${}^{\,+10.5}_{\,-10.3}$ to ${}^{\,+9.6}_{\,-9.9}$~pb
at $\sqrt s=8$~TeV. Again, a larger effect on the central value is
observed in the approach 
of~\cite{Cacciari:2011hy,Czakon:2011xx}:\footnote{When comparing the
  NNLO$_{\text{app}}$ results from {\prog} to those obtained with
  top++1.2~\cite{Czakon:2011xx} one finds a small difference of 
  $0.1$--$0.3$~pb. This is due
  to the implementation of the approximate NNLO corrections for the
  $gg$ initial state, where the exact tree cross sections for the
  separate colour-singlet and octet channels are used in {\prog},
  while the colour averaged tree cross sections are used in top++. In
  the threshold limit both implementations agree, so they are formally
  of the same accuracy.}
\[
\begin{array}{lllll}
158.7\,\mbox{pb} \;\mbox{($\!\!$\cite{Cacciari:2011hy})}
 & \;\to\;& 
156.2\,\mbox{pb} \;\mbox{($\!\!$\cite{Czakon:2011xx})} 
&\quad& \text{($7$~TeV)} ,\\[0.2cm]
226.6\,\mbox{pb} \;\mbox{($\!\!$\cite{Cacciari:2011hy})}
 & \;\to\;&
 222.5\,\mbox{pb}
 \;\mbox{($\!\!$\cite{Czakon:2011xx})}
 &\quad& \text{($8$~TeV)} .
\end{array} 
\]
While these results agree with ours within the quoted uncertainties,
the relative difference is larger than for the Tevatron which
indicates the relevance of the uncalculated fixed-order NNLO terms 
for the dominant gluon-fusion channel.

%%%%%%%%%%%%%%%%%%%%%%%%%%%%%%%%%%%%%%%%%%%%%%%%%%%%%%%%%%%%%%%%%
\section{Dependence on PDF sets}
\label{sec:PDF}

%%%%%%%%%%%%%%%%%%%%%%
\begin{table}[t]
  \caption{\label{tab::bcut}\sf Values for $\beta_{\text{cut}}$. The strong
    coupling constant is $\alpha_s(M_Z)=0.118$.
}
\vspace*{-0.0cm}
 \begin{center}
    \begin{tabular}{l|ccc|ccc|ccc}
      & \multicolumn{3}{|c|}{Tevatron} & \multicolumn{3}{|c|}{LHC7} &
      \multicolumn{3}{|c}{LHC8} \\
      PDF set & $k_s=1$ & $2$ & $4$ & $k_s=1$ & $2$ &
      $4$ & $k_s=1$ & $2$ & $4$ \\
      \hline
      MSTW2008   & 0.56 & 0.38\phantom{4} & 0.34 & 0.74 & 0.54\phantom{4} & 0.33 & 0.76 & 0.54\phantom{4} & 0.33 \\
      NNPDF2.1 & 0.53 & 0.29\phantom{4} & 0.19 & 0.74 & 0.51\phantom{4} & 0.25 & 0.75 & 0.52\phantom{4} & 0.25 \\
      ABM11    & 0.54 & 0.36\phantom{4} & 0.22 & 0.73 & 0.54\phantom{4} & 0.33 & 0.74 & 0.54\phantom{4} & 0.33 \\
      CT10     & 0.57 & 0.39\phantom{4} & 0.24 & 0.74 & 0.54\phantom{4} & 0.33 & 0.75 & 0.54\phantom{4} & 0.33
    \end{tabular}
  \end{center}
\vspace*{-0.2cm}
\end{table}
%%%%%%%%%%%%%%%%%%%%%%%%%%

In this section we present results for the cross section computed with
different PDF sets. As representatives we choose 
MSTW2008~\cite{Martin:2009iq},
NNPDF2.1~\cite{Ball:2011mu}, ABM11~\cite{Alekhin:2012ig}, and
CT10~\cite{Lai:2010vv}. (Note that CT10 at the moment provides only NLO PDFs.) 
We use $m_t=173.3$~GeV, and choose PDF sets for the common value 
$\alpha_s(M_Z)=0.118$ of the strong coupling constant. 
This provides complementary information to~\cite{Alekhin:2012ig}, 
where results for different PDF sets with their best, but often
rather different $\alpha_s$ values have been presented, which, 
due to the dependence of the partonic cross sections on $\alpha_s$, 
obscures the difference due to the parton densities themselves.

For NNPDF2.1 the total number of replicas used is $N_{{\rm rep}}=100$.
To determine the PDF$+\alpha_s$ error, this number is distributed over
sets with different $\alpha_s$ values according to a Gaussian
distribution with a standard deviation of $\delta\alpha_s=0.0015$.
For MSTW2008 and CT10 we provide asymmetric PDF$+\alpha_s$ errors using
the methods from \cite{Martin:2009bu,Lai:2010nw}. In the latter case
we use $\delta\alpha_s=0.002$ as the standard deviation.
For ABM11 
we compute symmetric errors with the formula
\begin{equation}
  \delta\sigma = \sqrt{\,\sum_k \left( \sigma_0 - \sigma_k \right)^2}\,,
\end{equation}
where the sum runs over all eigenvectors of the given set. This is the
same prescription as in~\cite{Alekhin:2012ig}. Note that in the case
of ABM11 $\alpha_s$ is one of the parameters of the fit. Thus, the PDF
error computed above already includes the uncertainty due to the
variation of the strong coupling constant.
For the NNLL calculation we also have to determine the parameter
$\beta_{\text{cut}}$. The values are given in Table~\ref{tab::bcut}.

%%%%%%%%%%%%%%%%%%%%%%%%%%%%
\begin{table}[!t]
\renewcommand{\tabcolsep}{0.8pc} % enlarge column spacing
\renewcommand{\arraystretch}{1.0} % enlarge line spacing
\caption{\label{tab:xs_all}\sf Total cross sections in pb at the
Tevatron and LHC ($\sqrt{s}=7\,$TeV and $8\,$TeV for different PDF 
sets and common strong coupling input $\alpha_s(M_Z)=0.118$.
See caption of Table~\ref{tab:tevatron173.3} 
for the definition of the various errors.}
\begin{center}
\begin{tabular}{llll}
\hline\\[-4.3mm]  
{\sf \hspace{-0.2cm} 
Tevatron} &${\text{NLO}}$&${\text{NNLO}}$&${\text{NNLL}}$
\\\hline\\[-3mm] 
     MSTW2008
     & $\phantom{0}{{6.48^{\,+0.33+0.23}_{\,-0.70-0.22}}}$
     & $\phantom{0}{{7.11^{\,+0.22+0.03+0.29}_{\,-0.32-0.03-0.24}}}$
     & $\phantom{0}{{7.26^{\,+0.22+0.03+0.30}_{\,-0.21-0.03-0.25}}}$
      \vspace{2mm} \\
     NNPDF2.1
     & $\phantom{0}{{6.77^{\,+0.34+0.23}_{\,-0.62-0.23}}}$
     & $\phantom{0}{{7.18^{\,+0.22+0.02+0.26}_{\,-0.17-0.02-0.26}}}$
     & $\phantom{0}{{7.49^{\,+0.26+0.02+0.27}_{\,-0.25-0.02-0.27}}}$
      \vspace{2mm} \\
     ABM11
     & $\phantom{0}{{6.45^{\,+0.26+0.14}_{\,-0.64-0.14}}}$
     & $\phantom{0}{{7.29^{\,+0.22+0.01+0.15}_{\,-0.31-0.01-0.15}}}$
     & $\phantom{0}{{7.46^{\,+0.22+0.01+0.16}_{\,-0.20-0.01-0.16}}}$
      \vspace{2mm} \\
     CT10
     & $\phantom{0}{{6.75^{\,+0.35+0.46}_{\,-0.74-0.35}}}$
     & $\phantom{0}{{7.33^{\,+0.22+0.03+0.52}_{\,-0.33-0.03-0.40}}}$
     & $\phantom{0}{{7.47^{\,+0.23+0.03+0.52}_{\,-0.22-0.03-0.41}}}$ \\[0.4cm]
\hline \\[-4.3mm]  
{\sf \hspace{-0.2cm} 
LHC7} &${\text{NLO}}$&${\text{NNLO}_{\text{app}}}$&${\text{NNLL$_{2}$}}$
\\\hline\\[-3mm] 
     MSTW2008
     & $\phantom{0}{{151.8^{\,+18.2+6.8}_{\,-19.9-6.2}}}$
     & $\phantom{0}{{164.3^{\,+10.6+4.4+7.2}_{\,-11.1-4.4-6.7}}}$
     & $\phantom{0}{{165.9^{\,+5.3+4.4+7.3}_{\,-5.6-4.4-6.8}}}$
      \vspace{2mm} \\
     NNPDF2.1
     & $\phantom{0}{{157.0^{\,+19.1+6.8}_{\,-19.7-6.8}}}$
     & $\phantom{0}{{161.9^{\,+8.3+4.3+7.1}_{\,-7.5-4.3-7.1}}}$
     & $\phantom{0}{{166.1^{\,+7.0+4.3+7.3}_{\,-8.3-4.3-7.3}}}$
      \vspace{2mm} \\
     ABM11
     & $\phantom{0}{{123.3^{\,+14.5+5.0}_{\,-16.1-5.0}}}$
     & $\phantom{0}{{146.7^{\,+9.4+3.8+5.8}_{\,-9.7-3.8-5.8}}}$
     & $\phantom{0}{{148.2^{\,+5.3+3.8+5.9}_{\,-5.3-3.8-5.9}}}$
      \vspace{2mm} \\
     CT10
     & $\phantom{0}{{148.7^{\,+17.7+12.7}_{\,-19.3-11.4}}}$
     & $\phantom{0}{{160.7^{\,+10.4+4.4+13.6}_{\,-10.8-4.4-12.2}}}$
     & $\phantom{0}{{162.3^{\,+5.2+4.4+13.7}_{\,-5.4-4.4-12.3}}}$\\[0.4cm]
\hline  \\[-4.3mm] 
{\sf \hspace{-0.2cm} 
LHC8} &${\text{NLO}}$&${\text{NNLO}_{\text{app}}}$&${\text{NNLL$_{2}$}}$
\\\hline\\[-3mm] 
     MSTW2008
     & $\phantom{0}{{217.5^{\,+26.0+9.2}_{\,-27.9-8.3}}}$
     & $\phantom{0}{{234.4^{\,+15.7+6.5+9.7}_{\,-16.0-6.5-9.0}}}$
     & $\phantom{0}{{236.5^{\,+7.4+6.5+9.8}_{\,-7.3-6.5-9.1}}}$
      \vspace{2mm} \\
     NNPDF2.1
     & $\phantom{0}{{225.1^{\,+27.3+9.0}_{\,-27.6-9.0}}}$
     & $\phantom{0}{{232.2^{\,+12.2+6.5+9.4}_{\,-11.0-6.5-9.4}}}$
     & $\phantom{0}{{238.0^{\,+9.7+6.5+9.7}_{\,-11.7-6.5-9.7}}}$
      \vspace{2mm} \\
     ABM11
     & $\phantom{0}{{180.8^{\,+21.4+6.8}_{\,-23.3-6.8}}}$
     & $\phantom{0}{{212.1^{\,+14.2+5.7+7.8}_{\,-14.3-5.7-7.8}}}$
     & $\phantom{0}{{214.1^{\,+7.4+5.7+7.9}_{\,-7.2-5.7-7.9}}}$
      \vspace{2mm} \\
     CT10
     & $\phantom{0}{{212.8^{\,+25.3+16.5}_{\,-27.1-14.8}}}$
     & $\phantom{0}{{229.2^{\,+15.5+6.4+17.6}_{\,-15.6-6.4-15.8}}}$
     & $\phantom{0}{{231.2^{\,+7.2+6.4+17.7}_{\,-7.2-6.4-15.9}}}$
\end{tabular}
\end{center}
\end{table}

The results for the cross section at the Tevatron and at the LHC with
$\sqrt{s}=7$~TeV and $\sqrt{s}=8$~TeV are given in
Table~\ref{tab:xs_all}.\footnote{The small difference in the MSTW2008 values 
with respect to Section~\ref{sec:nnlo} is due to
the different value of $\alpha_s(M_Z)$.}
We find good agreement between the different PDF sets for the cross
section at the Tevatron. At the LHC, we find again agreement 
between MSTW2008, NNPDF2.1, 
and CT10, but the value of ABM11 is 
significantly lower than the others. This reflects the smaller 
gluon density in the ABM11 fit that does not make use of the 
Tevatron jet data, which drive the high-$x$ gluon distribution 
upwards. The difference in the top cross section is significantly larger 
than the error assigned to the individual PDF fits, which indicates 
that it may be possible to use the top-pair production cross section 
to constrain the gluon density. We pursue this further in
Section~\ref{sec:gluon}.

%%%%%%%%%%%%%%%%%%%%%%%%%%%%%%%%%%%%%%%%%%%%%%%%%%%%%%%%%%%%%%%%%
\section{Heavy ``top'' quarks}
\label{sec:heavytop}

Heavy colour-triplet fermions appear in the extension of the standard model 
by a fourth generation, and are also often introduced  to cancel the 
top-loop contributions to the Higgs mass in non-supersymmetric models that 
aim at  addressing the hierarchy problem (such as Little Higgs 
models~\cite{ArkaniHamed:2002qy,Perelstein:2003wd}). In this context 
the extra fermions can be singlets under the $SU(2)_L$ gauge group or 
may transform under non-chiral representations, so they are less 
constrained than fourth generation quarks. Heavy colour-triplet fermions 
in other  $SU(2)_L$ representations with exotic electric charges have 
also been introduced in models of electroweak symmetry 
breaking~\cite{Agashe:2006at}. Within QCD, the production cross 
sections of these postulated particles only depend on the spin and 
colour charge and can be computed with {\prog}.

The LHC experiments have already performed direct searches for the 
production of heavy quarks in various decay channels, with the most 
stringent bounds  $m_B>611\,$GeV (for a heavy bottom quark $B$ decaying 
exclusively as $B\to tW$, using $4.7$~fb$^{-1}$  of 
data~\cite{Chatrchyan:2012yea}),  $m_T>557\,$GeV (for a heavy top quark
 decaying exclusively as  $T\to bW$, using $5.0$ fb$^{-1}$ of 
data~\cite{CMS:2012ab}),  $m_T>475\,$GeV  (for the decay $T\to tZ$ allowed 
for $SU(2)$ singlet $T$, using  $1.14$~fb$^{-1}$~\cite{Chatrchyan:2011ay}).

%%%%%%%%%%%%%%%%%%%%%%%%%%%%%%%%%%%%%%%%%%%%%%%%%%%%%%%%%%%%%%%%%%%%%%%%%%
\begin{table}[p]
\caption{\sf Total cross sections for the production of heavy quarks 
$Q$ beyond the standard model in pb at the LHC ($\sqrt{s}=7\,$TeV, 
upper table and $\sqrt{s}=8\,$TeV, lower table).
See caption of Table~\ref{tab:tevatron173.3} 
for the definition of the various errors. Heavy ``top'' mass $m_Q$ in GeV.}
\label{tab:mT-lhc78}
\newcommand{\m}{\hphantom{$-$}}
\newcommand{\cc}[1]{\multicolumn{1}{c}{#1}}
\renewcommand{\arraystretch}{1.0} % enlarge line spacing
\begin{center}
\begin{tabular}{llll}
\multicolumn{2}{l}{\tt LHC, $\sqrt{s}=7\,$TeV}&&\\[0.1cm]
  \hline\\[-4.3mm]  
  $m_Q$  &${\text{NLO}}$&${\text{NNLO}_{\text{app}}}$&${\text{NNLL$_{2}$}}$
  \\\hline\\[-3mm] 
  350
  & $\phantom{0}{3.11_{-0.40}^{+0.32}{}_{-0.15}^{+0.17}}$
  & $\phantom{0}{3.17_{-0.16}^{+0.13}{}_{-0.05}^{+0.05}{}_{-0.16}^{+0.18}}$
  & $\phantom{0}{3.21_{-0.11}^{+0.12}{}_{-0.05}^{+0.05}{}_{-0.16}^{+0.18}}$
  \vspace{2mm} \\
  400
  & $\phantom{0}{1.37_{-0.17}^{+0.13}{}_{-0.07}^{+0.08}}$
  & $\phantom{0}{1.39_{-0.07}^{+0.05}{}_{-0.02}^{+0.02}{}_{-0.07}^{+0.08}}$
  & $\phantom{0}{1.41_{-0.05}^{+0.06}{}_{-0.02}^{+0.02}{}_{-0.07}^{+0.08}}$
  \vspace{2mm} \\
  450
  & $\phantom{0}{6.44_{-0.81}^{+0.62}{}_{-0.32}^{+0.40}}\times 10^{-1}$
  & $\phantom{0}{6.56_{-0.30}^{+0.23}{}_{-0.09}^{+0.09}{}_{-0.34}^{+0.40}}\times 10^{-1}$
  & $\phantom{0}{6.66_{-0.24}^{+0.27}{}_{-0.09}^{+0.09}{}_{-0.34}^{+0.41}}\times 10^{-1}$
  \vspace{2mm} \\
  500
  & $\phantom{0}{3.20_{-0.40}^{+0.30}{}_{-0.16}^{+0.21}}\times 10^{-1}$
  & $\phantom{0}{3.26_{-0.14}^{+0.11}{}_{-0.04}^{+0.04}{}_ {-0.17}^{+0.21}}\times 10^{-1}$
  & $\phantom{0}{3.32_ {-0.11}^{+0.14}{}_{-0.04}^{+0.04}{}_{-0.17}^{+0.21}}\times 10^{-1}$
  \vspace{2mm} \\
  550
  & $\phantom{0}{1.67_ {-0.21}^{+0.15}{}_{-0.09}^{+0.11}}\times 10^{-1}$
  & $\phantom{0}{1.70_{-0.07}^{+0.05}{}_{-0.02}^{+0.02}{}_{-0.09}^{+0.11}}\times 10^{-1}$
  & $\phantom{0}{1.73_{-0.06}^{+0.07}{}_{-0.02}^{+0.02}{}_{-0.09}^{+0.11}}\times 10^{-1}$
  \vspace{2mm} \\
  600
  & $\phantom{0}{8.97_{-1.12}^{+0.82}{}_{-0.49}^{+0.63}}\times 10^{-2}$
  & $\phantom{0}{9.14_{-0.38}^{+0.26}{}_{-0.09}^{+0.09}{}_{-0.48}^{+0.63}}\times 10^{-2}$
  & $\phantom{0}{9.31_{-0.32}^{+0.40}{}_{-0.09}^{+0.09}{}_{-0.50}^{+0.64}}\times 10^{-2}$
  \vspace{2mm} \\
  650
  & $\phantom{0}{4.96_ {-0.62}^{+0.45}{}_{-0.28}^{+0.36}}\times 10^{-2}$
  &  $\phantom{0}{5.06_{-0.20}^{+0.14}{}_{-0.04}^{+0.04}{}_{-0.27}^{+0.37}}\times 10^{-2}$
  & $\phantom{0}{5.17_{-0.18}^{+0.23}{}_{-0.04}^{+0.04}{}_{-0.28}^{+0.38}}\times 10^{-2}$
  \vspace{2mm} \\
  700
  &  $\phantom{0}{2.81_{-0.35}^{+0.25}{}_{-0.17}^{+0.22}}\times 10^{-2}$
  &  $\phantom{0}{2.87_{-0.11}^{+0.08}{}_{-0.02}^{+0.02}{}_{-0.16}^{+0.22}}\times 10^{-2}$
  & $\phantom{0}{2.94_{-0.10}^{+0.13}{}_{-0.02}^{+0.02}{}_{-0.16}^{+0.23}}\times 10^{-2}$\\[1mm]
\hline\\[2.5mm]
\multicolumn{2}{l}{\tt LHC, $\sqrt{s}=8\,$TeV}&&\\[0.1cm]
\hline  \\[-4.3mm]
$m_Q$
  &${\text{NLO}}$&${\text{NNLO}_{\text{app}}}$&${\text{NNLL$_{2}$}}$
\\\hline\\[-3mm] 
      350
     & $\phantom{0}{{4.93_ {-0.62}^{+0.50}{}_{-0.23}^{+0.26}}}$
     & $\phantom{0}{5.03_{-0.26}^{+0.22}{}_ {-0.09}^{+0.09}{}_{-0.24}^{+0.27}}$
     & $\phantom{0}{5.09_ {-0.18}^{+0.19}{}_{-0.09}^{+0.09}{}_{-0.24}^{+0.28}}$
      \vspace{2mm} \\
      400
     & $\phantom{0}{{2.23^{\,+0.22+0.12}_{\,-0.28-0.11}}}$
     & $\phantom{0}{2.27_{-0.11}^{+0.09}{}_{-0.04}^{+0.04}{}_{-0.11}^{+0.13}}$
     & $\phantom{0}{2.30_{-0.08}^{+0.09}{}_{-0.04}^{+0.04}{}_{-0.11}^{+0.13}}$
      \vspace{2mm} \\
      450
     & $\phantom{0}{1.08_ {-0.13}^{+0.10}{}_ {-0.05}^{+0.06}}$
     & $\phantom{0}{1.10_ {-0.05}^{+0.04}{}_{-0.02}^{+0.02}{}_{-0.05}^{+0.06}}$
     & $\phantom{0}{1.12_ {-0.04}^{+0.04}{}_{-0.02}^{+0.02}{}_{-0.06}^{+0.06}}$
      \vspace{2mm} \\
      500
     & $\phantom{0}{5.55_ {-0.68}^{+0.51}{}_ {-0.27}^{+0.33}}\times 10^{-1}$
     & $\phantom{0}{5.65_ {-0.25}^{+0.19}{}_{-0.07}^{+0.07}{}_{-0.28}^{+0.34}}\times 10^{-1}$
     & $\phantom{0}{5.73_ {-0.20}^{+0.23}{}_{-0.07}^{+0.07}{}_{-0.29}^{+0.34}}\times 10^{-1}$
      \vspace{2mm} \\
      550
     & $\phantom{0}{2.97_{-0.36}^{+0.27}{}_{-0.15}^{+0.19}}\times 10^{-1}$
     & $\phantom{0}{3.03_ {-0.13}^{+0.10}{}_{-0.04}^{+0.04}{}_{-0.15}^{+0.19}}\times 10^{-1}$
     & $\phantom{0}{3.07_ {-0.11}^{+0.13}{}_ {-0.04}^{+0.04}{}_ {-0.16}^{+0.19}}\times 10^{-1}$
      \vspace{2mm} \\
      600
     & $\phantom{0}{1.65_ {-0.20}^{+0.15}{}_{-0.08}^{+0.11}}\times 10^{-1}$
     & $\phantom{0}{1.68_ {-0.07}^{+0.05}{}_{-0.02}^{+0.02}{}_{-0.09}^{+0.11}}\times 10^{-1}$
     & $\phantom{0}{1.71_{-0.06}^{+0.07}{}_ {-0.02}^{+0.02}{}_{-0.09}^{+0.11}}\times 10^{-1}$
     \vspace{2mm} \\
      650
     & $\phantom{0}{9.44_ {-1.15}^{+084}{}_{-0.50}^{+0.63}}\times 10^{-2}$
     & $\phantom{0}{9.62_{-0.39}^{+0.27}{}_{-0.09}^{+0.09}{}_{-0.49}^{+0.64}}\times 10^{-2}$
     & $\phantom{0}{9.78_{-0.33}^{+0.41}{}_{-0.09}^{+0.09}{}_{-0.51}^{+0.65}}\times 10^{-2}$
     \vspace{2mm} \\
      700
     & $\phantom{0}{5.54_{-0.68}^{+0.49}{}_{-0.30}^{+0.39}}\times 10^{-2}$
     & $\phantom{0}{5.64_ {-0.22}^{+0.16}{}_{-0.05}^{+0.05}{}_ {-0.29}^{+0.39}}\times 10^{-2}$
     & $\phantom{0}{5.74_{-0.19}^{+0.25}{}_{-0.05}^{+0.05}{}_{-0.30}^{+0.40}}\times 10^{-2}$
 \vspace{2mm} \\
      750
     & $\phantom{0}{3.31_ {-0.41}^{+0.29}{}_{-0.19}^{+0.24}}\times 10^{-2}$
     & $\phantom{0}{3.38_ {-0.13}^{+0.09}{}_{-0.03}^{+0.03}{}_{-0.18}^{+0.25}}\times 10^{-2}$
     & $\phantom{0}{3.44_{-0.12}^{+0.15}{}_{-0.03}^{+0.03}{}_{-0.18}^{+0.25}}\times 10^{-2}$
\vspace{2mm} \\
      800
     & $\phantom{0}{2.02_ {-0.25}^{+0.18}{}_{-0.12}^{+0.15}}\times 10^{-2}$
     & $\phantom{0}{2.06_ {-0.08}^{+0.05}{}_{-0.02}^{+0.02}{}_{-0.11}^{+0.16}}\times 10^{-2}$
     & $\phantom{0}{2.10_ {-0.07}^{+0.09}{}_{-0.02}^{+0.02}{}_ {-0.11}^{+0.16}}\times 10^{-2}$\vspace{2mm} \\
      850
     & $\phantom{0}{1.25_ {-0.15}^{+0.11}{}_ {-0.08}^{+0.10}}\times 10^{-2}$
     & $\phantom{0}{1.27_ {-0.05}^{+0.03}{}_{-0.01}^{+0.01}{}_{-0.07}^{+0.10}}\times 10^{-2}$
     & $\phantom{0}{1.30_ {-0.04}^{+0.06}{}_{-0.01}^{+0.01}{}_{-0.07}^{+0.11}}\times 10^{-2}$\vspace{2mm} \\
      900
     & $\phantom{0}{7.81_{-0.97}^{+0.69}{}_{-0.50}^{+0.65}}\times 10^{-3}$
     &$\phantom{0}{7.99_ {-0.29}^{+0.20}{}_{-0.05}^{+0.05}{}_{-0.46}^{+0.69}}\times 10^{-3}$
     & $\phantom{0}{8.18_{-0.26}^{+0.37}{}_{-0.05}^{+0.05}{}_{-0.47}^{+0.70}}\times 10^{-3}$ \vspace{2mm} \\ 
     950
     & $\phantom{0}{4.94_ {-0.61}^{+0.44}{}_{-0.33}^{+0.43}}\times 10^{-3}$
     &$\phantom{0}{5.07_ {-0.18}^{+0.13}{}_ {-0.03}^{+0.03}{}_ {-0.30}^{+0.46}}\times 10^{-3}$
     &  $\phantom{0}{5.19_ {-0.16}^{+0.23}{}_{-0.03}^{+0.03}{}_{-0.31}^{+0.47}}\times 10^{-3}$ \vspace{2mm} \\
      1000
     &$ \phantom{0}{3.15_{-0.40}^{+0.28}{}_ {-0.22}^{+0.29}}\times 10^{-3}$
     &$\phantom{0}{3.24_ {-0.12}^{+0.08}{}_{-0.02}^{+0.02}{}_{-0.20}^{+0.31}}\times 10^{-3}$
     & $\phantom{0}{3.32_{-0.11}^{+0.15}{}_{-0.02}^{+0.02}{}_{-0.21}^{+0.32}}\times 10^{-3}$ \\[1mm]
\hline
\end{tabular}\\[2pt]
\end{center}
\end{table} 

In Table~\ref{tab:mT-lhc78} we present our predictions 
for the  production cross section of heavy quarks $Q$ at the LHC with $7$~TeV 
centre-of-mass energy for the mass range $m_Q=350$--$700$~GeV that is relevant 
for the analysis of $5$~fb$^{-1}$ of data collected in the 2011 run, and 
for masses up to $1$~TeV at the LHC operating at  $\sqrt s=8$~TeV.
For each mass the value of the parameter $\beta_{\text cut}$ appearing in 
the momentum-space resummation formalism with a running soft scale has 
been determined using the procedure  introduced in~\cite{Beneke:2011mq}. 
The resulting values range from $\beta_{\text cut}=0.50$--$0.42$ for 
$m_Q=350$--$700$~GeV and $\sqrt s=7$~TeV, and from 
$\beta_{\text cut}=0.51$--$0.39$ for $m_Q=350$--$1000$~GeV at the LHC with 
$\sqrt s=8$~TeV, and are implemented as default values in \prog.
We use the MSTW2008 PDFs and the running strong coupling with 
five active quark flavours and therefore neglect the impact of the top 
quark on the evolution of these quantities. The partonic cross sections 
are also computed with five massless flavours, which corresponds to 
replacing the top quark $t$ by $Q$ (rather than adding $Q$ to the known 
quarks).
The same approximation was also used in~\cite{Cacciari:2011hy}.

For the mass range  $m_Q=350$--$700$~GeV at the LHC with $\sqrt s=7$~TeV, 
the NNLL corrections grow from about $3\%$ to above $4\%$ relative to 
the NLO cross section evaluated with the appropriate NLO PDFs, while 
for $m_Q=1$~TeV at the LHC with $\sqrt s=8$~TeV the NNLL corrections become 
larger than $5\%$. In this comparison, the higher-order corrections are 
partially compensated by the change from NLO to NNLO PDFs, with the 
associated change in the value of the strong coupling. 
Normalizing the higher-order results instead to the NLO cross section 
evaluated with MSTW2008NNLO PDFs, the NNLL corrections range from 
$10\%$ for $m_Q=350$~GeV to $14\%$ for $m_Q=1000$~GeV at the LHC with $\sqrt s=8$~TeV, 
while the approximate NNLO cross section amounts to a $9$--$11\%$ 
correction. Therefore higher-order effects beyond NNLO are small, 
but reach a magnitude of the order of the residual uncertainty of 
the approximate NNLO result for larger masses. 

The total theoretical uncertainty defined
below~\eqref{bestTevatron}
is largely independent of the mass and centre-of-mass
energy. At $\sqrt{s}=8$~TeV, in the mass range
$m_Q=350$-$1000$~GeV, it ranges from
${}_{-13\%}^{+10\%}$ to ${}_{-13\%}^{+9\%}$ at NLO,
from ${}^{+5\%}_{-5\%}$ to ${}^{+3\%}_{-4\%}$ at NNLO, and from
${}^{+4\%}_{-4\%}$ to ${}^{+5\%}_{-3\%}$ at NNLL.  It is counterintuitive
that for larger masses the error estimate of the approximate NNLO 
prediction becomes smaller than that for NNLL. However, the size of our
uncertainty estimate for large masses is similar to that found
in~\cite{Cacciari:2011hy}, so this issue is common to the Mellin-space
and momentum space results. 
The relative PDF uncertainty is similar at NLO, NNLO$_{\text{app}}$ 
and NNLL and 
grows from ${}^{+5\%}_{-5\%}$ to $ {}^{+10\%}_{-6\%}$ for 
$m_Q=350$--$1000$~GeV at $\sqrt s=8$~TeV.
Performing the computation
with the NNPDF2.1 PDF set we find that the results for NNLL$_2$ 
(NNLO$_{\text{app}}$) are
smaller by 4--6\% (3\%) if the same coupling constant is used 
for both PDF sets. Therefore
both sets agree within the estimated PDF uncertainty.

\begin{figure}[t]
  \begin{center}
    \includegraphics[width=0.65\textwidth]{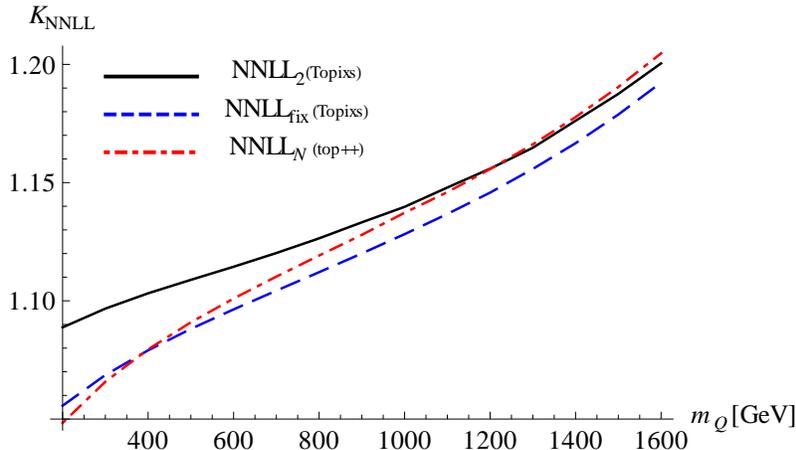}
    \caption{\label{fig:heavyNNLL}\sf Comparison of the ratio
      $K_{\text{NNLL}}=\sigma_{t\bar t}^{\text{NNLL}}/\sigma_{t\bar
        t}^{\text{NLO}}$ for heavy quark production at the LHC with
      $\sqrt s=8$~TeV for our default implementation NNLL$_2$ (black,
      solid) to our fixed-scale implementation NNLL$_{\text{fix}}$
      (blue, dashed) and the results NNLL$_N$ (red, dot-dashed) from
      the Mellin-space approach~\cite{Cacciari:2011hy,Czakon:2011xx}. The NNLL results are normalized by the NLO cross section computed with NNLO PDFs.}
  \end{center}
\end{figure}

Since the contribution of the threshold region to the total cross
section is more important for larger masses, we expect that the
ambiguities in the implementation of threshold resummation are reduced
and the predictions of different formalisms show better agreement than
for lighter quarks. We therefore compare our default implementation,
NNLL$_2$, to an implementation NNLL$_{\text{fix}}$ of resummation in
momentum space with a fixed soft scale, determined using the method
proposed in~\cite{Becher:2007ty}, that was denoted as ``Method 1''
in~\cite{Beneke:2011mq}. This implementation is available in
{\prog} with the option \texttt{MUSRUN=0}, and the default values of the
fixed soft scale are implemented for MSTW2008 with $m_Q=173.3$--$2000$~GeV.
By NNLL$_N$ we denote the results obtained using the program
top++1.2~\cite{Czakon:2011xx} that implements resummation using the
traditional Mellin-moment method~\cite{Cacciari:2011hy}.  In
Figure~\ref{fig:heavyNNLL}  the results of the three implementations
are shown for the LHC with $8$~TeV centre-of-mass energy for the mass
range $m_Q=200$--$1600$~GeV. The larger masses are not relevant for
phenomenology with $20$~fb$^{-1}$ luminosity, but are included here in
order to compare the three implementations in a regime where the
threshold approximation is expected to be more accurate. We plot the
quantity $K_{\text{NNLL}}=\sigma_{t\bar t}^{\text{NNLL}}/\sigma_{t\bar
  t}^{\text{NLO}}$, where for comparison the NLO cross section in the
denominator is evaluated with the NNLO PDFs, in contrast to our
default treatment. It is seen that the three implementations are in
good agreement once the higher-order corrections become larger than
$10\%$, with larger differences for smaller masses. The difference of
the Mellin-space implementation to our default is always of a similar
magnitude or smaller than that of the two momentum-space
implementations\footnote{The results from {\prog} include bound-state
  effects and the resummation of Coulomb corrections beyond NNLO not
  included in top++. However, the bound-state contribution to
  $K_{\text{NNLL}}$ is of the order of $0.3$--$0.4$ percent for the
  mass range considered in Figure~\ref{fig:heavyNNLL}, so the conclusion of 
  the comparison to top++ is not affected. For top-pair production the
  contribution of the Coulomb corrections beyond NNLO above threshold
  was found to be even smaller~\cite{Beneke:2011mq}, which we also
  expect to be the case for larger masses.} so we conclude that the
momentum-space and Mellin-space approaches agree within the
ambiguities inherent in the threshold approximation.

%%%%%%%%%%%%%%%%%%%%%%%%%%%%%%%%%%%%%%%%%%%%%%%%%%%%%%%%%%%%%%%%%
\section{Gluon distribution function}
\label{sec:gluon}

The precise knowledge of parton distribution functions (PDFs) and of 
their uncertainties is a crucial requirement for the LHC physics program, 
particularly when the error on PDFs represents the largest single theoretical 
uncertainty, as is the case in top-pair production. PDFs are extracted 
by several collaborations~\cite{Martin:2009iq,Ball:2011mu,Alekhin:2012ig,JimenezDelgado:2009tv,Lai:2010vv,Aaron:2009aa} 
from a large number of measurements collected at 
different experiments, the early 
fixed-target deep-inelastic scattering (DIS) and Drell-Yan (DY) 
experiments and the modern HERA and Tevatron colliders. 
In the meantime the LHC, thanks to the accumulated high statistics and 
the well-controlled systematics, is providing the PDF fitting collaborations 
with a large amount of precise data. Soon the LHC measurements are going 
to provide essential constraints on most PDF combinations. 
The NNPDF collaboration has already published a PDF analysis, NNPDF2.2,  
which includes early electro-weak LHC data from the 2010 
runs~\cite{Ball:2011gg}. A recent study of the 
impact of LHC $W$ lepton charge-asymmetry data on a
MSTW2008 Monte Carlo parton set was presented in~\cite{Watt:2012tq}.

In this section we investigate the possibility to constrain the 
gluon distribution in the proton through the top-pair production cross 
section by combining the most accurate available theoretical calculation 
presented in this paper with LHC data. This is interesting in several 
respects. First, while the gluon distribution function 
at medium--small $x$ is rather 
well constrained by the precise HERA measurements of the scaling of 
the structure functions, at medium--large $x$ the gluon is currently 
constrained primarily from the jet inclusive measurements performed 
at the Tevatron. A more precise determination of the gluon at 
medium--large $x$, which is the region 
probed by the $t\bar{t}$ measurements at the LHC, would lead to 
an improved theoretical 
precision of the Higgs cross section in gluon fusion. Moreover 
the top-pair production cross section 
is correlated to the strong coupling constant and therefore 
potentially reduces the uncertainty 
in the determination of $\alpha_s$ from PDF global analyses.

In this section, we quantify the impact of the LHC measurements of 
the total $t\bar{t}$ cross section on the gluon uncertainty by using the 
re-weighting technique~\cite{Ball:2010gb,Ball:2011gg}. This technique is
valid for Monte Carlo sets of PDFs, like those provided by the 
NNPDF collaboration. 
These sets determine a Monte Carlo representation of the probability 
density in the space of PDFs. This feature allows to include 
new experimental data by using Bayes' theorem, i.e. by re-weighting an 
existing PDF set (prior probability) without having to perform a new fit 
of the entire data.  The effect of a new independent dataset is 
quantified by computing for each replica $k$ 
($k=1,...,N_{\rm rep}$) of the starting PDF set a weight $w_k$, 
which assesses the
probability that the PDF replica agrees with the new data. 
The re-weighted ensemble then forms a representation of the probability 
distribution of PDFs conditional on both the old and the new data. 
The weights are computed straightforwardly by 
evaluating the $\chi^2_k$ of the new data to each of the replicas, 
according to Eq.~(9) of~\cite{Ball:2010gb}.
The distribution can then be unweighted and a new PDF set, 
which now includes both the data fitted in the initial PDF set and 
the additional data, is extracted. 

Before presenting the results of this analysis, it is important to 
specify the three ingredients that are employed in order to determine 
the $\chi^2_k$ of the LHC top-pair production data to each replica 
and therefore the weights $w_k$: the experimental 
data, the theoretical predictions and the prior PDF set.

The (single) experimental data added to the PDF analysis is obtained
from the combination of the ATLAS and CMS $t\bar t$ cross section
measurements presented in~\cite{ATLAS-CONF-2011-121,ATLAS:2012aa,ATLAS-CONF-2011-140,CMS-PAS-TOP-11-024}.
While the proper combination of these results can only be performed by
the experimental collaborations, we simply combine them here 
in a single data point by adding in quadrature statistical and
systematic errors of the ATLAS and CMS combined results, and by adding
the luminosity errors as a 100\% correlated uncertainty between the
two experiments, which results in
\begin{equation}
\label{sigttforfit}
\sigma^{\rm exp.}_{t\bar{t}}=173.23 \,\pm\, 6.55\, {\rm (stat. + syst.)}\,\pm\, 7.00\, {\rm (lumi.)}\,\, {\rm pb} = 173.23\ \pm 9.59 \,{\rm pb}. 
\end{equation}
The theoretical predictions $\sigma^{\rm th.}_{t\bar{t},k}$ are
obtained for each PDF replica $k$ from~\eqref{eq:match2}.\footnote{The
  present analysis has been performed
  before~\cite{Baernreuther:2012ws} appeared and hence does not
  include the full $q\bar{q}$ NNLO partonic cross section. However, as
  discussed above, this has a negligible effect on the central value
  of the gluon-dominated $t\bar t$ cross section at LHC.  Since the
  present re-weighting analysis does not include theoretical
  errors, only the central value matters.}  Since the PDF errors do
not account for theoretical uncertainties, the latter are neither
included in the computation of the weights $w_k$.  As starting PDF
sets, we consider two cases.  In the first part of the analysis, the
prior probability is given by the NNPDF2.1 NNLO parton set. The latter
is a NNLO determination of parton distributions from a global set of
hard-scattering data, including DIS, Drell-Yan and jet data. Each set
is determined for a given value of $\alpha_s(M_Z)$, which is treated
as an external parameter.  Several sets are available with
$\alpha_s^j=0.114,0.115,...,0.124$.  Since the $t\bar{t}$ cross
section is strongly correlated to $\alpha_s$, which in turn is
strongly correlated to the gluon density, it is important to take into
account the $\alpha_s$ uncertainty in the prior PDF set.  The
$N_{\alpha_s}^j$ replicas determining the starting parton set (such
that $\sum_{j=1}^{N_{\alpha_s}}\,N_{\alpha_s}^j = 350$) are taken from
various $\alpha_s^j$ sets, as it was done in Section~\ref{sec:PDF} to
determine the PDF$+\alpha_s$ uncertainty, according to a Gaussian
distribution centered on $\alpha_s(M_Z)=0.118$ with a standard
deviation of $\delta\alpha_s$= 0.0015.

\begin{figure}[t]
\centering
\includegraphics[width=0.48\textwidth]{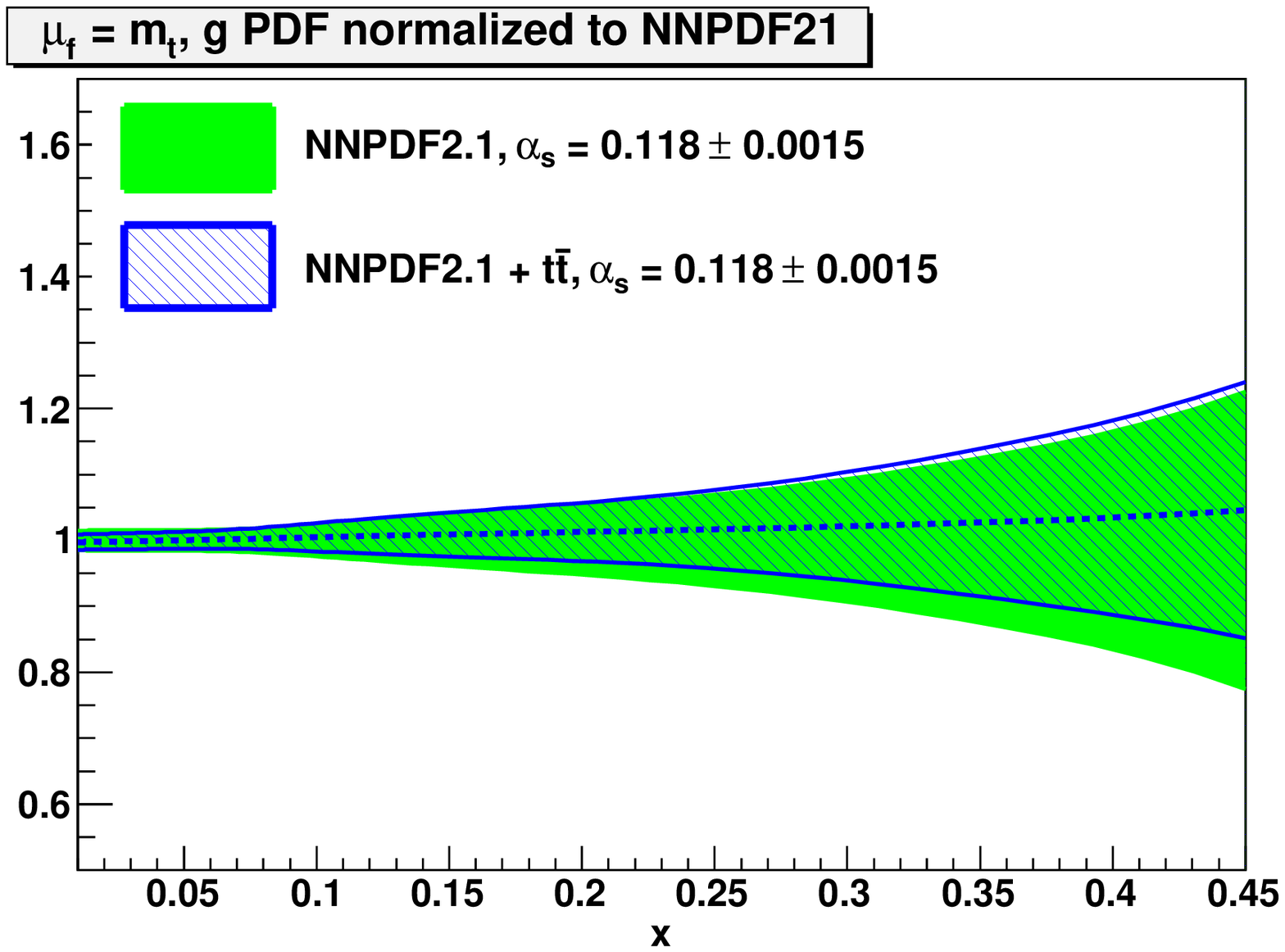}
\includegraphics[width=0.48\textwidth]{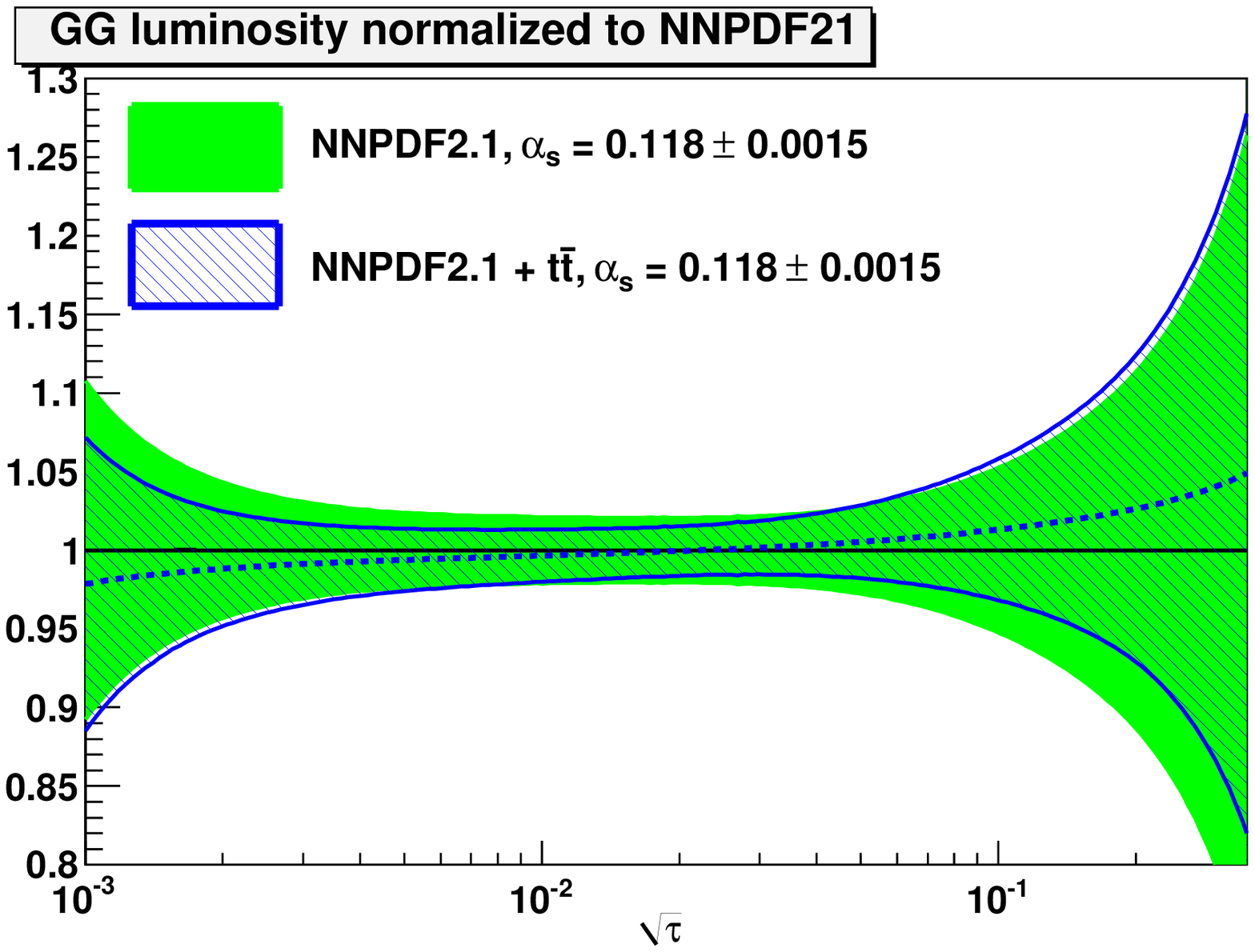}\\
\includegraphics[width=0.48\textwidth]{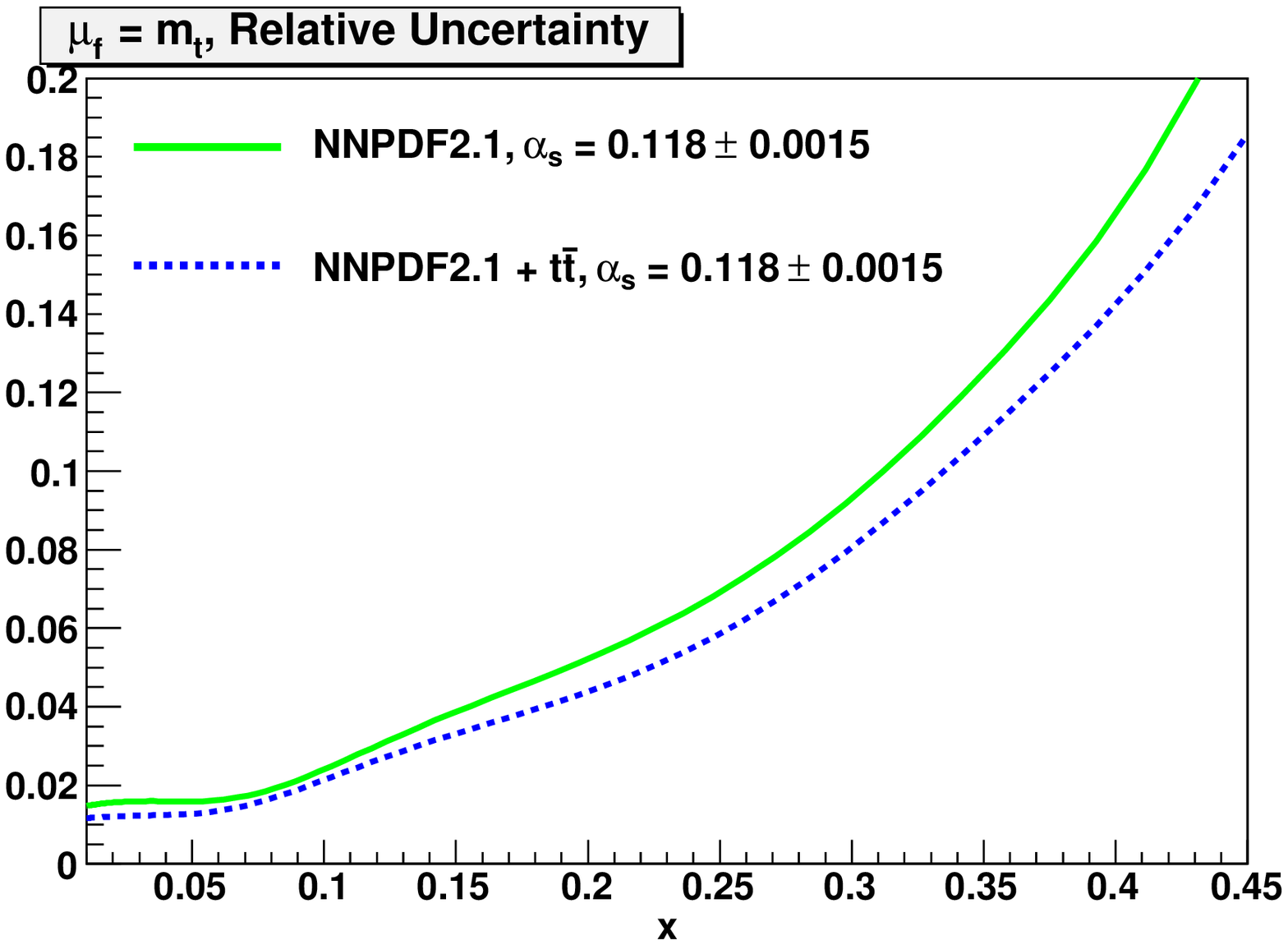}
\includegraphics[width=0.48\textwidth]{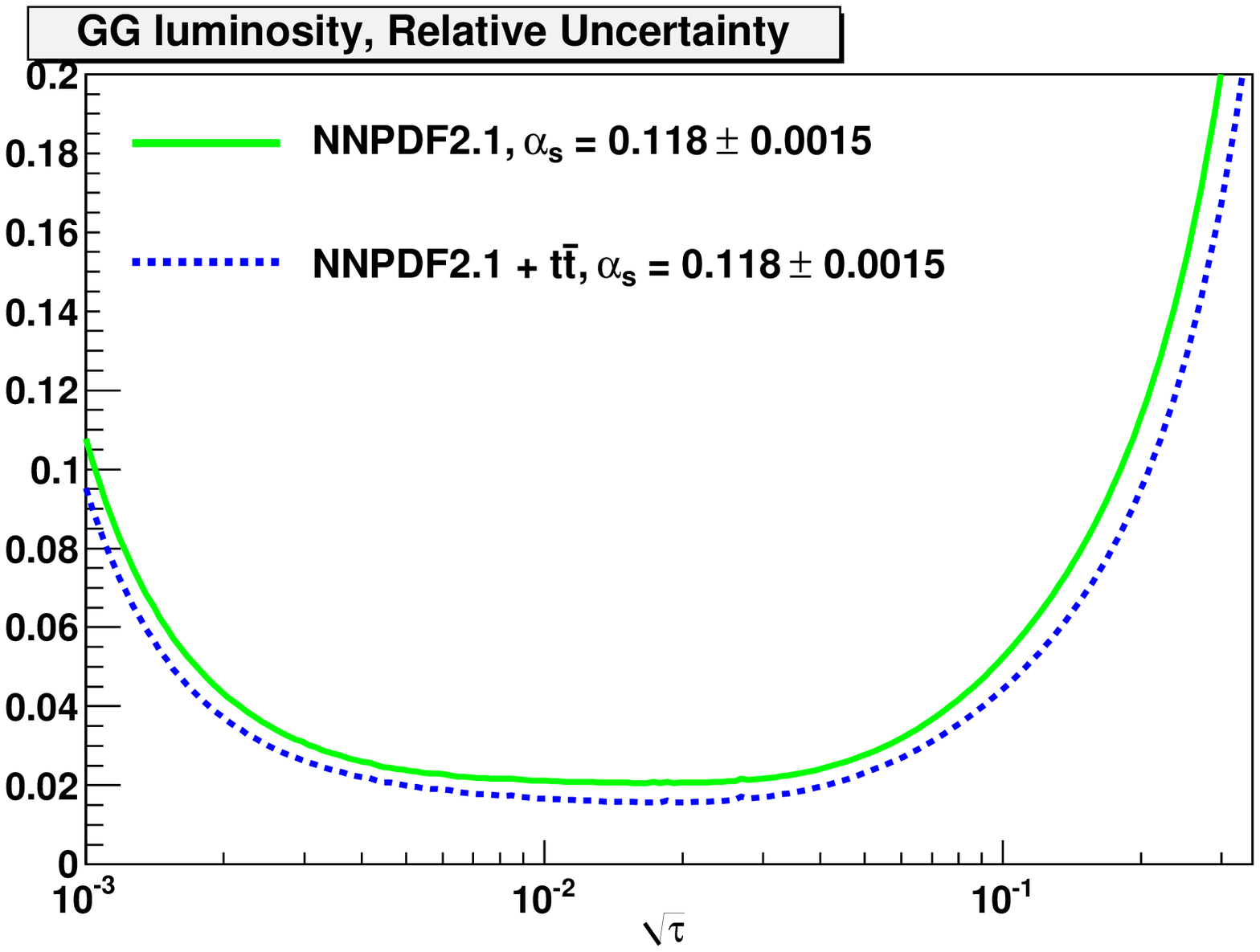}
\caption{\label{fig:asvar-global}\sf Effect of including 
the total $t\bar{t}$ cross-section measurement into 
the NNPDF2.1 global analysis for varying $\alpha_s$. Top: 
Gluon distribution (left) 
and $gg$ luminosity (right) before (solid, solid green in color) and 
after (shaded, shaded blue in color) including the $t\bar{t}$ 
measurement in the NNPDF fit. Both curves are normalized to the central value of the NNPDF2.1 
gluon before re-weighting. Bottom: Half-width of the error bands 
in the upper plots 
before (solid, solid green in color) and after (dashed, dashed blue in color) 
the inclusion of the $t\bar{t}$ measurement in the fit.}
\end{figure}

The results of the re-weighting are displayed in 
Figure~\ref{fig:asvar-global}. On the left-hand side the impact 
of including the LHC measurement on the gluon density, with 
$\mu_f=m_t=173.3\,$GeV, is shown. 
As expected, there is an effect in the large-$x$ region. The 
uncertainty in the gluon density given by the width of the bands 
can be assessed directly from the bottom-left plot, where the 
two lines refer to before and after including the $t\bar t$ 
cross section in the PDF analysis. Since the $t\bar t$ cross section 
predicted with the NNPDF2.1 PDF is close to (\ref{sigttforfit}), 
see Table~\ref{tab:xs_all}, there is only a small effect on the 
central gluon distribution. However, the uncertainty is reduced 
by about 20\% for $x\gtrsim 0.15$. The same results are shown 
on the right-hand side of Figure~\ref{fig:asvar-global} but for 
the gluon-gluon luminosity, defined by 
\begin{equation}
 \mathcal{L}_{gg}(\tau)=\frac{1}{s}
\displaystyle\int^1_\tau \frac{dy}{y} \ f_{g/p}(y,\mu_f) \ 
f_{g/p}(\tau/y,\mu_f),
\end{equation}
where $s$ is the LHC centre-of-mass energy squared and 
$\mu$, following Ref.~\cite{Campbell:2006wx}, is set to $\mu_f=\sqrt{\tau s}$.
A similar impact as the one observed on the gluon density is 
observed on the $gg$ luminosity in the expected region 
$\sqrt{\tau} \ge 2m_t/\sqrt{s}\,\sim\, 0.05$ 
and partially also in the smaller $\sqrt{\tau}$ region due to the 
momentum conservation sum rules. 
The predicted $t\bar{t}$ cross section evaluated with the NNPDF2.1 set
changes from 
$\sigma^{\rm th}_{t\bar{t}}=166.1 \,\pm\, 7.0 \,{\rm (PDF)}\,{\rm pb}$ 
to $\sigma^{\rm th}_{t\bar{t}}=168.4 \,\pm\, 5.4 \,{\rm (PDF)}$ after 
re-weighting; the central value gets closer to the 
experimental data, and the PDF uncertainty is reduced by about 25\%.
On top of that, we observe that the $\chi^2$ per degree of freedom to the 
Tevatron jet data is reduced by the inclusion of the $t\bar{t}$ measurement.
If the same analysis were performed by using as a prior set 
the NNPDF2.1 set with fixed $\alpha_s$, the impact would be rather smaller, 
as the reduction of the uncertainty is not only due to the gluon, 
but also to $\alpha_s(M_Z)$, whose value computed over the replicas increases 
from $0.1180\pm 0.0015$ to $0.1184\pm 0.0013$ after including  
the $t\bar{t}$ data in the fit. 

\begin{figure}[t]
\centering
\includegraphics[width=0.48\textwidth]{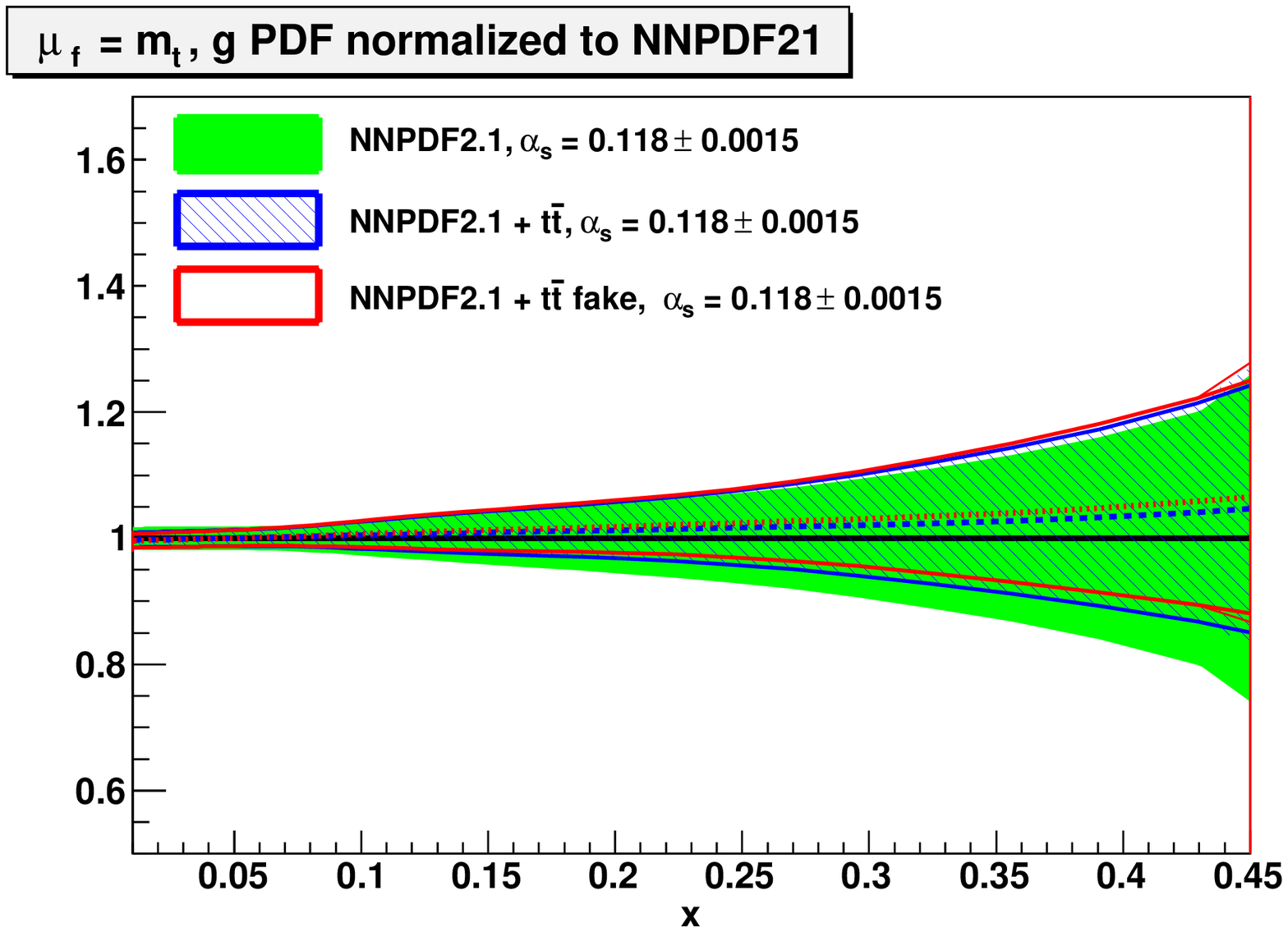}
\includegraphics[width=0.48\textwidth]{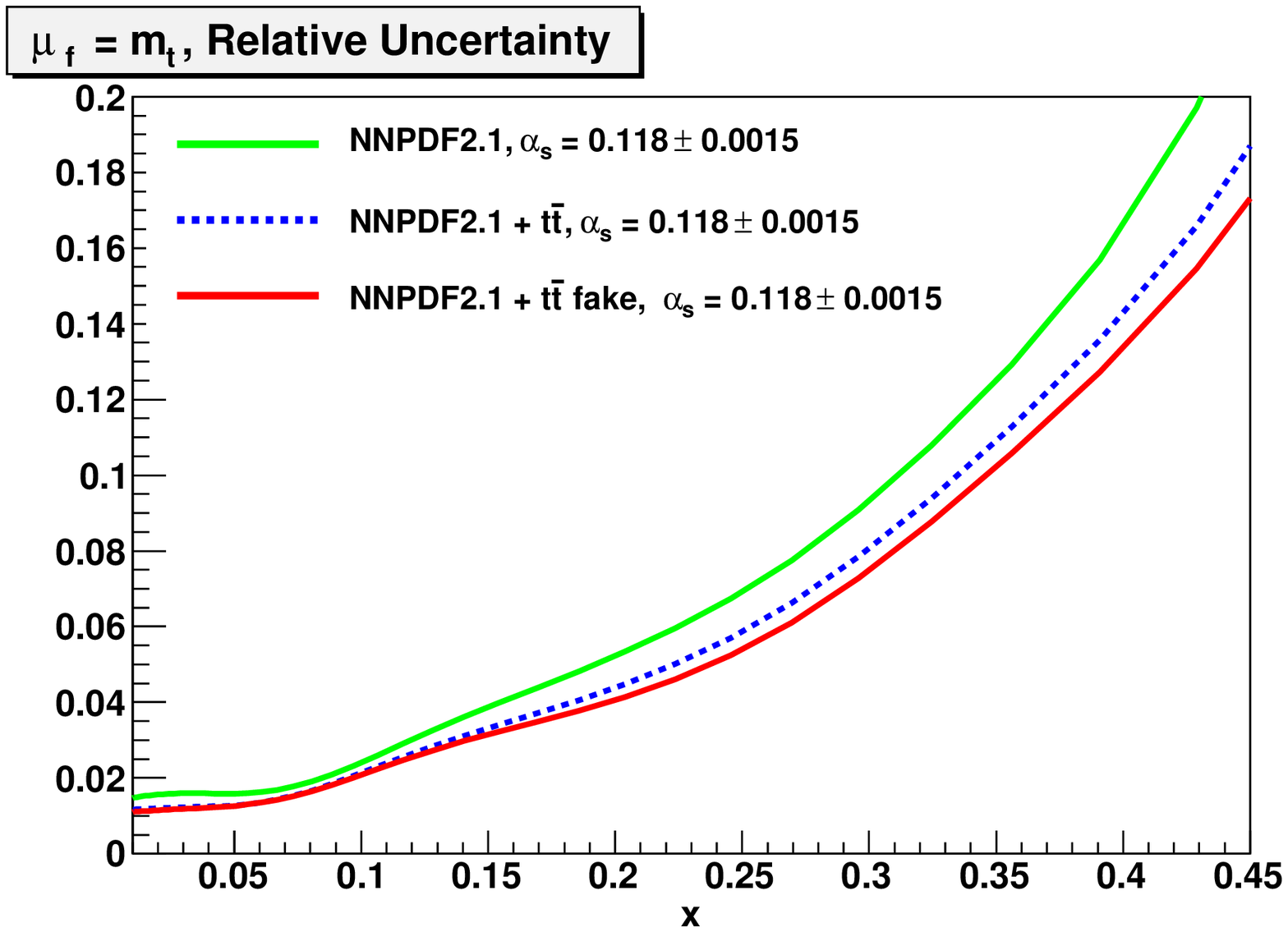}
\caption{\label{fig:asvar-global-ideal}\sf Effect of including  
the total $t\bar{t}$ cross section measurement with reduced uncertainty 
into the NNPDF2.1 global analysis for varying $\alpha_s$. 
Gluon distribution (left) and (right) half-width 
of the gluon error band  in the left plot before 
(solid light gray, solid green in color) and after  
including the $t\bar{t}$ measurement in the fit. In this case the 
dashed curve (dashed, dashed blue in color) refers to the inclusion
of the real measurement (\ref{sigttforfit}), 
the solid dark gray curve (solid, solid red in color) refers to the 
inclusion of the measurement with reduced uncertainty (\ref{fakeXS}).}
\end{figure}

It is interesting to observe that the effect is already visible with 
the data collected in the 2011 7 TeV LHC run. We can then expect an 
even more significant effect, when in the near future the experimental 
precision will reach the present theoretical accuracy. We therefore 
show in Figure~\ref{fig:asvar-global-ideal} what happens if we suppose 
that the measured $t\bar{t}$ cross section had an error as large as 
the theoretical uncertainty,\footnote{Here we adopt the error from the 
MSTW2008 PDF set in Table~\ref{tab:xs_all} that leads to a slightly 
smaller theoretical uncertainty.} while the central value 
is kept fixed,
\begin{equation}
\label{fakeXS}
\sigma^{\rm fake}_{t\bar{t}} = 173.23^{+6.7}_{-6.9} \,{\rm pb}, 
\end{equation}
and observe a further reduction by 5\% (to a total of 25\%) 
of the uncertainty in the gluon 
distribution to the NNPDF2.1 gluon in the large-$x$ 
region. This means that, on the one hand, the limiting factor will 
soon be the theoretical uncertainty, which will be reduced further 
only once the full NNLO  $gg$ partonic cross section is known. 
On the other hand, it is remarkable to see a significant effect 
from a single additional measurement that competes
with thousands of other measurements already included in PDF global 
analyses, in particular with the approximately 200 inclusive 
jet data points collected at Tevatron Run II that are 
relevant in the large-$x$ region. A larger effect should therefore 
be expected, once $t\bar{t}$ rapidity 
and invariant mass distributions will be included in the PDF fits. 

In order to disentangle the effect of the jet data from the one of the 
top-pair production data, we now start from the NNPDF2.1\_DIS+DY 
set as prior probability, 
which has the same features as the NNPDF2.1 global set, but does 
not contain the jet data. This implies that the PDF uncertainty of the 
gluon at medium--large $x$ is larger. We can therefore assess 
whether the top cross section pushes the gluon into the same direction
as the Tevatron jet data and compare the constraining power of top-pair
production relative to the inclusive jet measurements. Furthermore, the 
starting PDF set is supposedly closer to the ABM11 NNLO PDF 
set~\cite{Alekhin:2012ig}, which does not include
the Tevatron data, allowing us to draw some conclusion on the differences 
observed in Section~\ref{sec:PDF} between theoretical
predictions obtained with ABM11 and the other PDF sets.

\begin{figure}[t]
\centering
\includegraphics[width=0.48\textwidth]{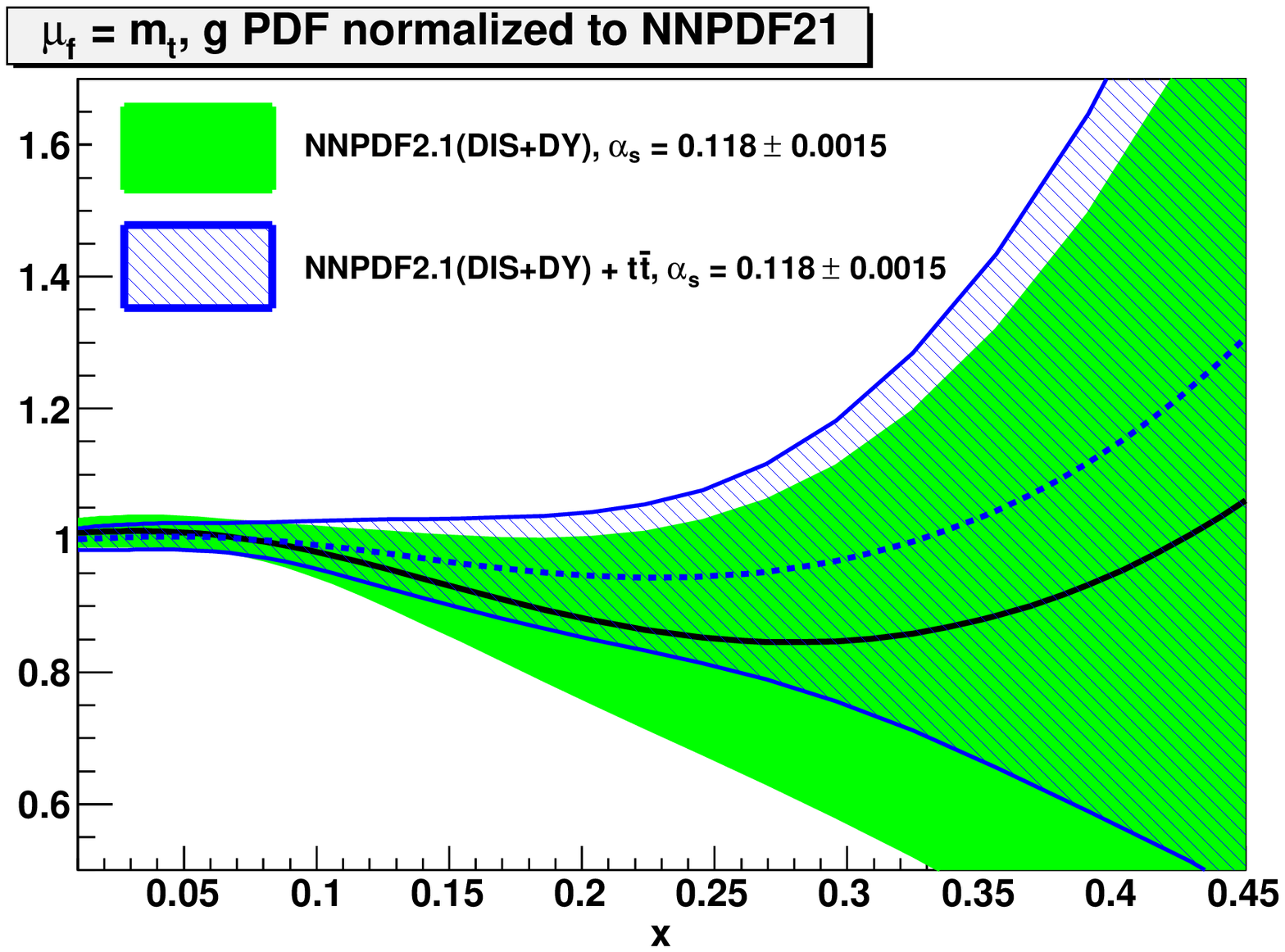}
\includegraphics[width=0.48\textwidth]{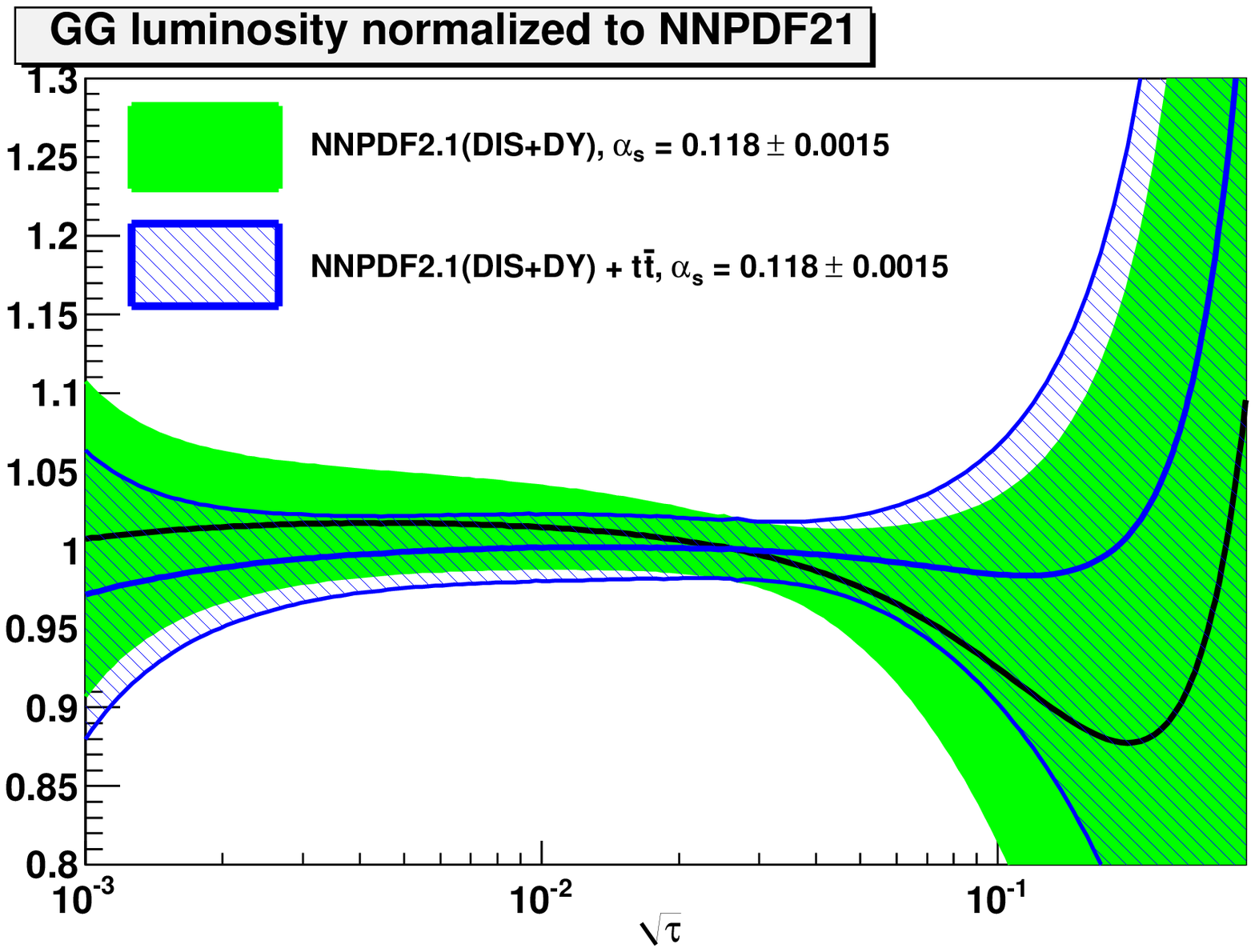}\\
\includegraphics[width=0.48\textwidth]{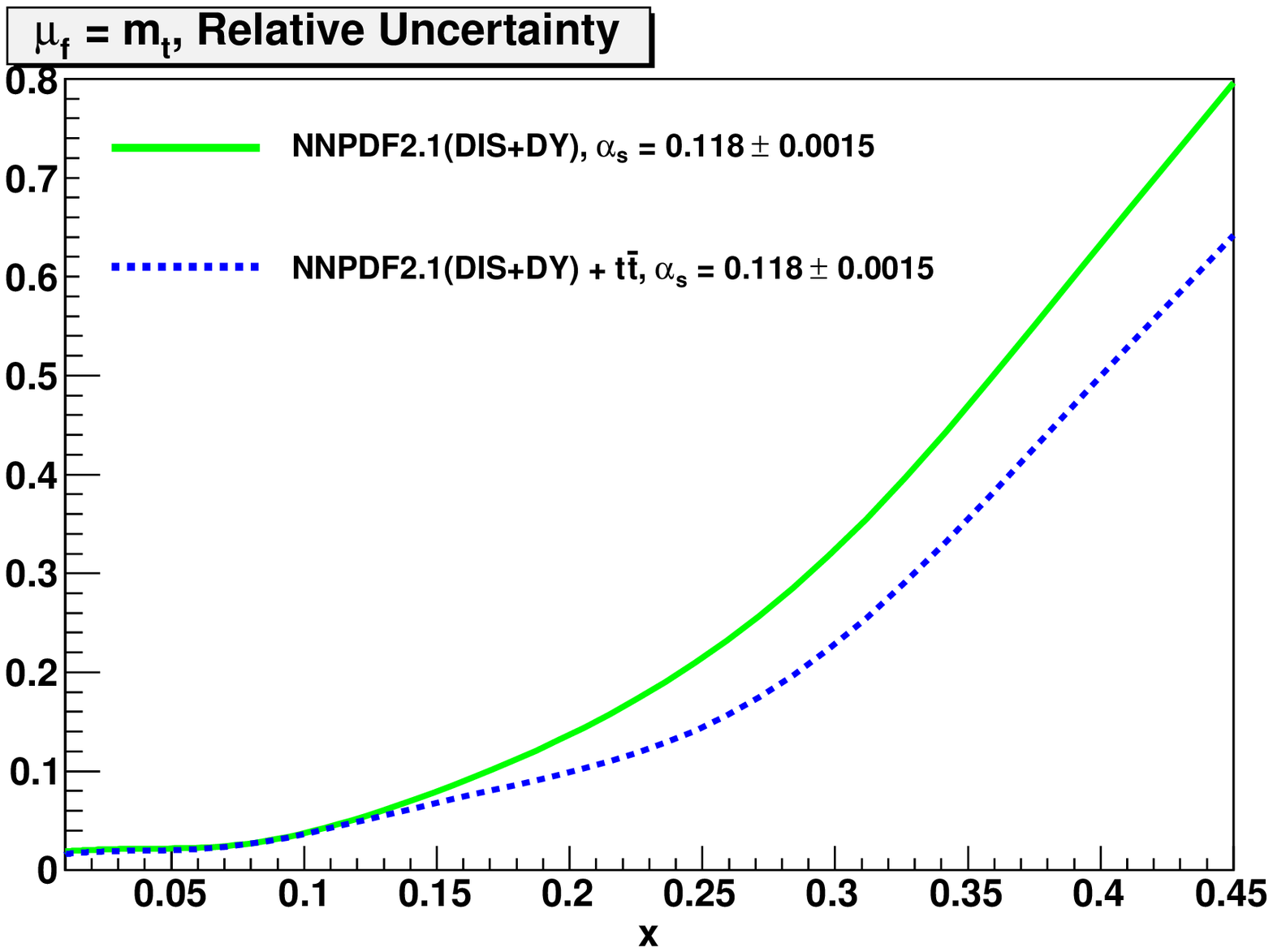}
\includegraphics[width=0.48\textwidth]{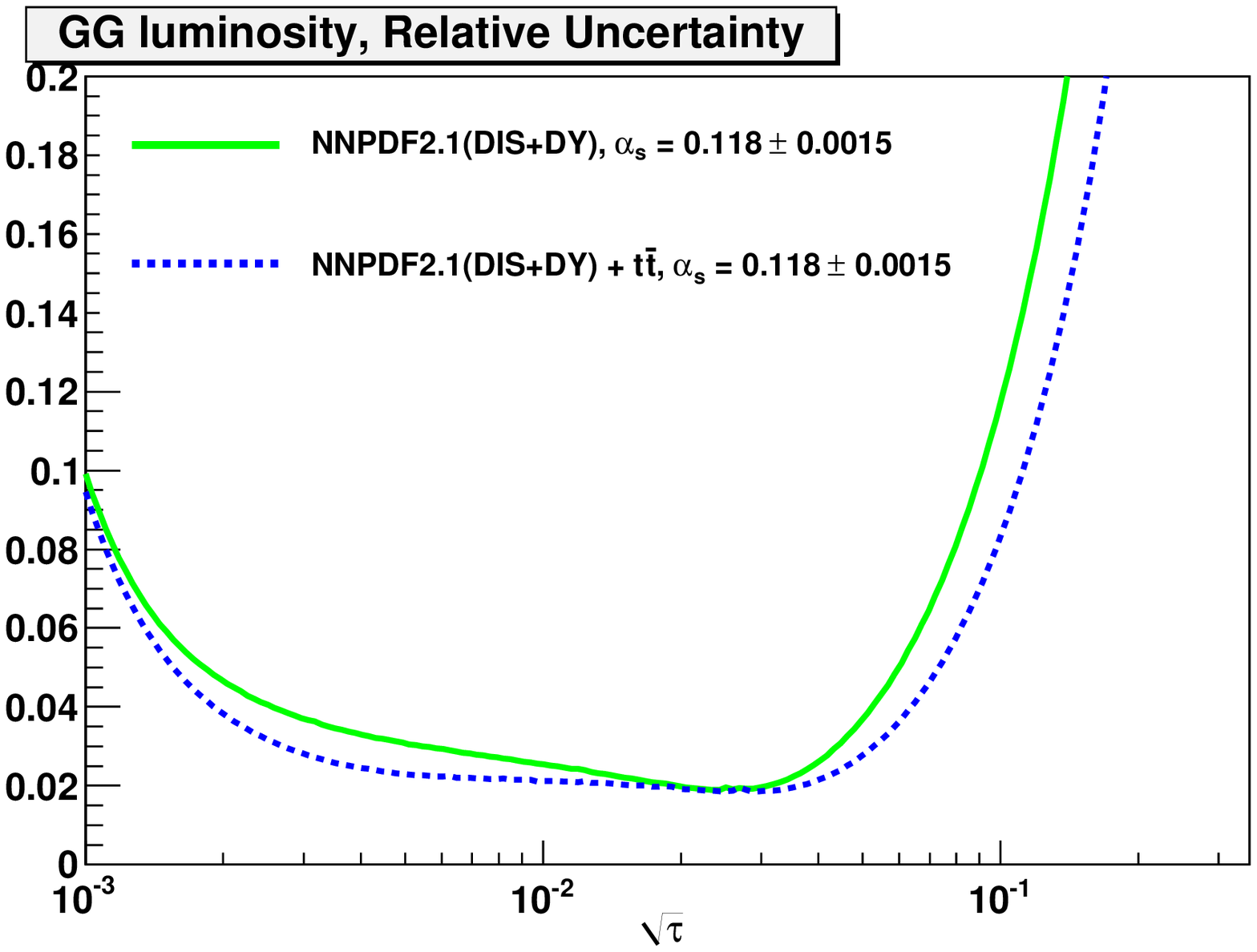}
\caption{\label{fig:asvar-nojets}\sf Effect of including the total 
$t\bar{t}$ cross section measurement (but excluding jet data)
on the NNPDF2.1\_DIS+DY PDF sets for varying $\alpha_s$. Top: 
Gluon distribution (left) and $gg$ luminosity (right) before 
(solid, solid green in color) 
and after (shaded, shaded blue in color) 
the inclusion of the $t\bar{t}$ measurement 
in the NNPDF fit. Both curves are normalized to the central value of the 
NNPDF2.1 gluon. Bottom: Half-width of the error 
bands before (solid, solid green in color) and after 
(dashed, dashed blue in color) including the $t\bar{t}$ measurement 
in the fit.}
\end{figure}

The theoretical prediction for the total top-pair production cross 
section obtained with the NNPDF2.1\_DIS+DY set
is $\sigma^{\rm th.}_{t\bar{t}}=161.1 \,\pm\, 9.9 \,{\rm (PDF)}\,{\rm pb}$,  
3\% smaller than the one obtained with the NNPDF2.1 global analysis, but 
compatible within the PDF uncertainty. 
The $\chi^2$ per degree of freedom of the $t\bar{t}$ data 
point to $\sigma^{\rm th.}_{t\bar{t}}$ is 1.6. 
After including the $t\bar{t}$ measurement in the fit the predicted 
cross section moves to 
$\sigma^{\rm th.}_{t\bar{t}}=167.5 \,\pm\, 6.8 \,{\rm (PDF)}\,{\rm pb} $ 
and the $\chi^2$ decreases to 0.4. 
In Figure~\ref{fig:asvar-nojets} we show the change of 
the gluon PDF and of the $gg$ luminosity due to the new fit.
The shift of the central value of the gluon is now significant, 
and it goes in the same direction as the shift when the 
jet data is included, bringing the black central line closer to one, which is 
the reference gluon of the global NNPDF2.1 fit. This can be 
expected since, as we observed earlier, the $\chi^2$ per degree of freedom
to the Tevatron jet data in the NNPDF2.1 global fit slightly improved
with the inclusion of the $t\bar{t}$ measurement in the fit.
A similar observation holds for the $gg$ luminosity. 
The error reduction is significant and amounts to about 35\% both for 
the gluon at large-$x$ and the gluon-gluon luminosity at medium--large 
$\sqrt{\tau}$. If we compare the size of the blue error band in 
Figure~\ref{fig:asvar-nojets}
to the one of the green error band in Figure~\ref{fig:asvar-global}, 
we see that the error reduction due to the inclusion 
of the jet data is certainly larger, by a factor 4, due to the larger 
number of data included. However, it is striking that the single 
$t\bar t$ data point already has a non-negligible effect  
with respect to the Tevatron jet data. 

\begin{figure}[t]
\centering
\includegraphics[width=0.48\textwidth]{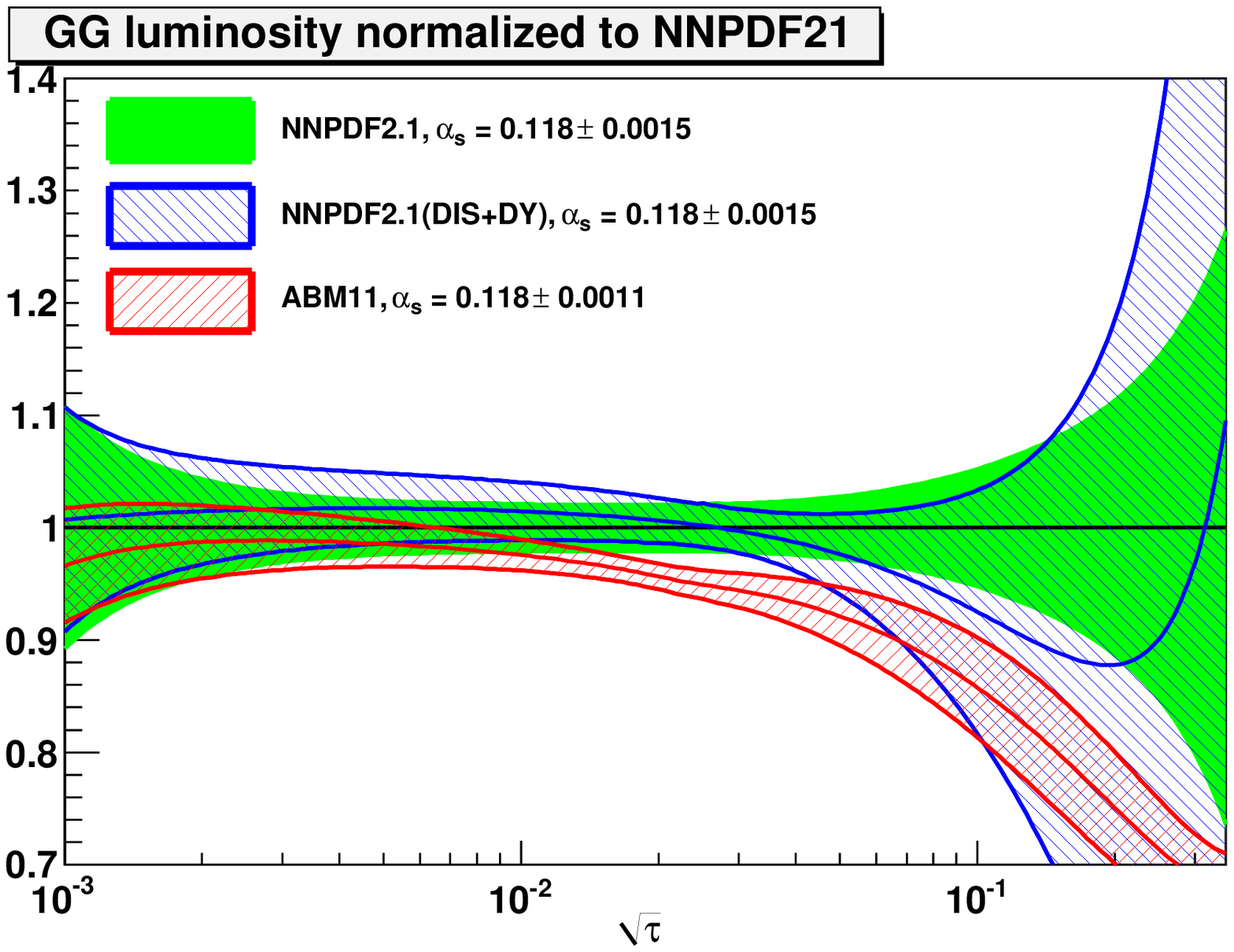}
\includegraphics[width=0.48\textwidth]{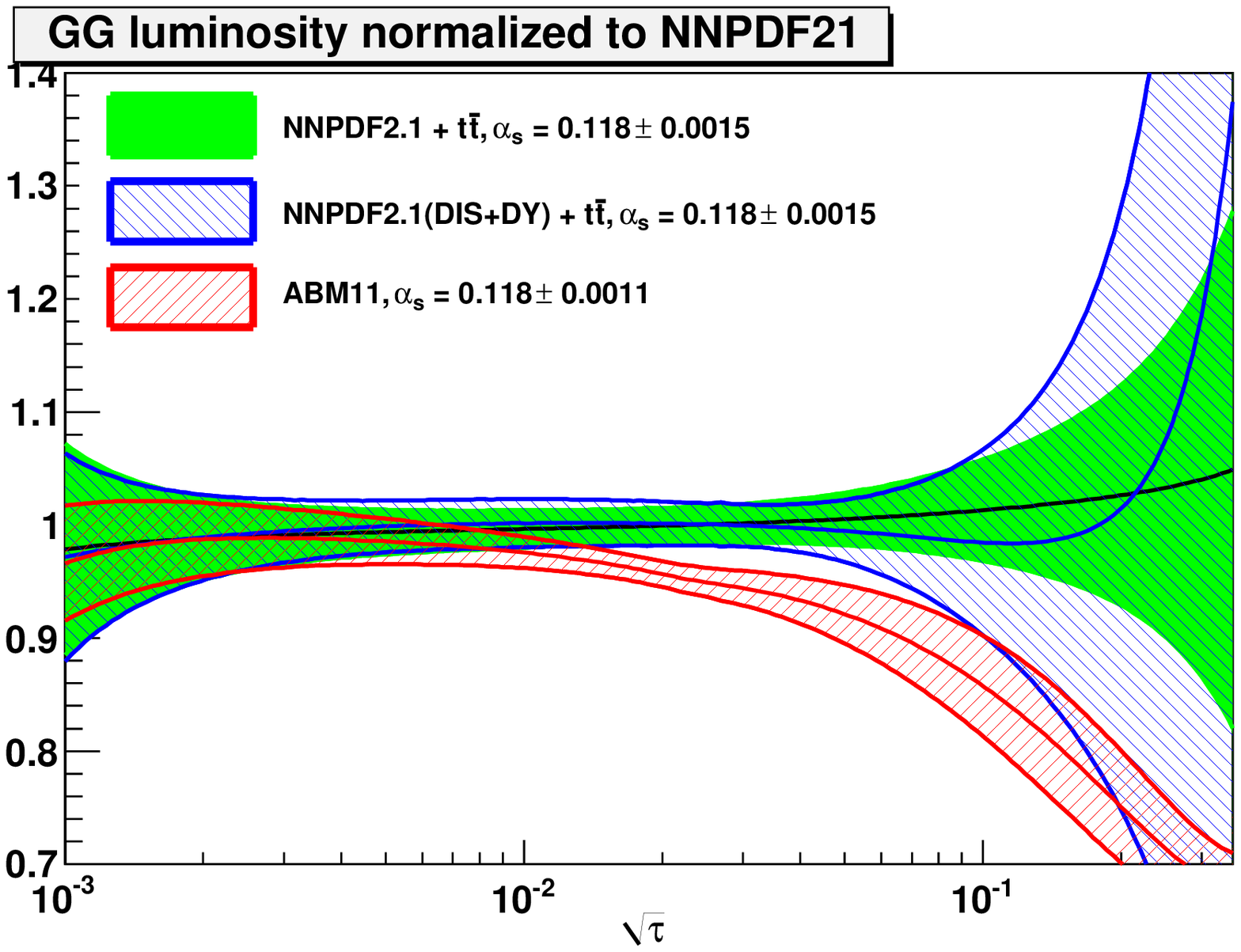}\\
\caption{\label{fig:abm}\sf Comparison of $gg$ luminosity obtained 
in the NNPDF2.1, NNPDF2.1\_DIS+DY 
and ABM11 parton fits before (left) and after (right) 
the $t\bar{t}$ measurement is included in the NNPDF2.1 and 
NNPDF2.1\_DIS+DY fits. 
All curves are normalized to the 
central value of the NNPDF2.1 gluon and computed for 
$\alpha_s(M_Z)=0.118$. }
\end{figure}

In Section~\ref{sec:PDF} we observed that the MSTW2008, NNPDF2.1, 
and CT10 PDF sets predict comparable values for the top-pair 
production cross sections 
at the LHC, while the ABM11 prediction is significantly lower than the 
others. On the left-hand side of Figure~\ref{fig:abm} we compare the ABM11 
gluon luminosity to the NNPDF2.1\_DIS+DY and 
NNPDF2.1 ones, using the same  $\alpha_s(M_Z)=0.118$ for consistency.
In the large-$x$ region the ABM11 gluon luminosity is about
$2\sigma$ smaller than NNPDF2.1. It is compatible with NNPDF2.1\_DIS+DY, 
mainly because the uncertainty of the latter is much larger, which 
in turn is mostly due to differences in the gluon parameterization, 
and partially also due to a different treatment of the 
higher-twist effects in DIS.  

On the right-hand side of Figure~\ref{fig:abm} we show the same 
results but after including the top-pair production cross section 
in the two NNPDF2.1 fits. We already noted that this increases 
the large-$x$ gluon and therefore the medium--large $\sqrt{\tau}$ 
gluon luminosity. A similar effect is expected on the ABM11
gluon and should bring it closer to the other global determinations. 
But, due to the small uncertainty of the ABM11 gluon, the effect would 
be much more significant, and the inclusion of the $t\bar t$ data 
might even require a more
flexible large-$x$ gluon parameterization.

A caveat to the analysis in the present section must be mentioned, 
namely that it was performed for a fixed top-quark mass value of 
$m_t=173.3\,$GeV. A change of $\pm 1$ GeV in the value of $m_t$, which 
corresponds to the present top-mass uncertainty, changes the 
predicted inclusive top production cross section by roughly $\pm 3\%$, 
which decreases the significance of this measurement for the 
NNPDF2.1 fit. On the other hand, the difference between the 
predicted $t\bar{t}$ cross section obtained with the NNPDF2.1 set, 
$\sigma^{\rm th.}_{t\bar{t}}=166.1 \,\pm\, 7.3 \,{\rm (PDF)}\,{\rm pb}$, 
and with the ABM11 set,  
$\sigma^{\rm th.}_{t\bar{t}}=148.2 \,\pm\, 5.9 \,{\rm (PDF)}\,{\rm pb}$
is about 10\%, see Table~\ref{tab:xs_all}. 
Hence, our conclusion that the present  
$t\bar{t}$ cross section measurement should already have an impact 
on the ABM11 fit remains valid after including the top-quark mass 
uncertainty.

%%%%%%%%%%%%%%%%%%%%%%%%%%%%%%%%%%%%%%%%%%%%%%%%%%%%%%%%%%%%%%%%%
\section{Conclusion}

In this paper we extended our analysis of the resummed
total top-quark pair production cross section \cite{Beneke:2011mq} in 
several directions. We updated our results for Tevatron, and
for LHC with 7 and 8 TeV centre-of-mass energy to account for the full 
$q\bar q$ NNLO partonic cross section~\cite{Baernreuther:2012ws}; 
studied the dependence on different parton-distribution inputs and the
impact of the LHC cross section measurement on the gluon distribution; 
and presented predictions for heavy ``top'' quarks. 
All results and many variations can be obtained with
the publicly available program \prog, which includes soft and Coulomb 
resummation. The main results of the present analysis can be summarized 
as follows:

\begin{itemize}
\item Including the full $q\bar q$ NNLO partonic cross section removes 
the slight tension that existed in the prediction of the $t\bar t$ 
cross section at the Tevatron from different groups in favour of the 
somewhat larger values and reduces the uncertainty by almost a
factor of two to about 3\%, see also~\cite{Baernreuther:2012ws}. 
Our NNLO+NNLL Tevatron $t\bar t$ production cross section result, 
assuming $m_t=173.3\,$GeV and $\alpha_s(M_Z)=0.1171$, is 
\begin{equation}
\sigma_{t\bar t} = 7.15^{\,+0.21}_{\,-0.20}\,\mbox{(theory)}
{}^{\,+0.30}_{\,-0.25} \,\mbox{(PDF+$\alpha_s$)} \,\mbox{pb}
\qquad (\mbox{MSTW2008NNLO}).
\end{equation}
The cross section increases to $7.26\,$pb,  
when the ``world average'' value $\alpha_s(M_Z)=0.1180$ is used.
A top mass  $m_t=171.4^{+5.4}_{-5.7}$~GeV is obtained from
Tevatron data.
\item 
Our prediction for the $t\bar t$ production cross section at 
the LHC with $\sqrt{s}=8\,$TeV, 
assuming $m_t=173.3\,$GeV and $\alpha_s(M_Z)=0.1171$, is 
\begin{equation}
\sigma_{t\bar t} =
231.8^{\,+9.6}_{\,-9.9}\,\mbox{(theory)} {}^{\,+9.8}_{\,-9.1}
\,\mbox{(PDF+$\alpha_s$)} \,\mbox{pb}
\qquad (\mbox{MSTW2008NNLO}).
\end{equation} 
The cross section increases to $236.5\,$pb,  
when the ``world average'' value $\alpha_s(M_Z)=0.1180$ is used.
\item The 2011 LHC data from the $\sqrt{s} = 7\,$ TeV run decrease 
the uncertainty of the gluon distribution at $x \gtrsim 0.15$ by about 20\%, 
using NNPDF2.1 as the reference gluon. This
effect is quite remarkable, given that only a single data point is 
added to the fit. The precision of the data and theoretical prediction 
will soon raise the issue of defining a
procedure for including consistently theoretical uncertainties into 
the PDF fits and their error estimates.
\item PDF fits including the jet data from the Tevatron agree very well 
in their predicted top cross section at the LHC. The PDF uncertainty is 
now the dominant theoretical uncertainty. The ABM11 fit predicts a 10\% 
lower cross section due to its smaller large-$x$ gluon distribution. 
Our study leads us to expect that the $t\bar t$ cross section
should already have a significant impact on this fit, even when 
all theoretical uncertainties,
including the top-quark mass input are accounted for.
\end{itemize}

\subsubsection*{Acknowledgements}
We would like to thank M. Czakon for discussions.
The work of M.B., J.P. and M.U. is supported by the DFG
Sonder\-for\-schungs\-bereich/Trans\-regio~9 ``Computergest\"utzte
Theoreti\-sche Teilchenphysik''.  P.F. acknowledges support 
by the ``Stichting voor Fundamenteel Onderzoek
der Materie (FOM)". This research was supported in part 
by the National Science Foundation under Grant 
No.~PHY05-51164.

\section*{Appendix: Functionality of 
T{\normalsize OPIXS}}
\label{sec:topixs}
In this section we describe the main functionality of the program
\prog. A detailed manual is available together with the program
package at the URL
\begin{center}
\verb+http://users.ph.tum.de/t31software/topixs/+
\end{center}
\prog\ uses {\sc hplog}~\cite{Gehrmann:2001pz} for the numerical
evaluation of harmonic polylogarithms and
{\sc quadpack}~\cite{quadpack} for numerical integrations. The PDF
sets are included via the {\sc lhapdf}~\cite{Whalley:2005nh}
library. The determination of the auxiliary parameter $\beta_{\text{cut}}$
and the fixed soft scale, 
which are used in the computation of the
resummed cross section, uses
routines from the GNU Scientific Library~\cite{GSL}. 

The program package consists of several {\sc fortran} and {\sc c++}
programs which implement the calculation of the total cross section as
well as that of the parameter $\beta_{\text{cut}}$ and  the fixed
soft scale.  The user interface consists
of a {\sc bash} shell script, called {\tt topixs}, which automatically
compiles and executes these programs. The user can change the program
settings by editing a single text file. Examples of such a
configuration file are included in the program package.

In our approach the cross section depends either on the auxiliary
parameter $\beta_{\text{cut}}$ or the fixed soft scale. While programs for
their computation are provided, we have also implemented default
values for several PDF sets as well as the values for heavy quarks
discussed in Section~\ref{sec:heavytop}. Therefore, it is usually not
necessary to compute them, in which case the GNU Scientific Library is
not required either.

Several approximations can be chosen for the cross-section
computation. A fixed-order calculation is possible to NLO or
NNLO$_{{\rm (app)}}$ accuracy. In the latter case, the user can choose
whether or not to include the full result for the $q\bar{q}$ channel
by changing the value of the variable {\tt QQNNLOEXACT} in the
configuration file. The resummed cross section can be evaluated in all
of the different approximations defined in \cite{Beneke:2011mq},
i.e. (N)NLL$_1$ and (N)NLL$_2$ with fixed soft scale, and (N)NLL$_2$
with running soft scale. The option {\tt BOUNDS} allows the user to
decide whether or not to include the contribution from bound states,
or even to compute only this contribution.

Our default choice is to use a running soft scale ({\tt MUSRUN=1}) and
include the bound-state contribution ({\tt BOUNDS=1}). The full result
for the $q\bar{q}$ channel is included in the NNLO computation and the
matching of NNLL$_2$ ({\tt QQNNLOEXACT=1}). These settings will be used
unless the variables are explicitly reset in the configuration file.

The main parameters of the calculation are the top-quark mass, the
collider type (proton-proton or proton-antiproton) and energy, and the
PDF set to be used. All of them can be chosen by the user. The default
settings are $m_t=173.3$~GeV ({\tt MTOP=173.3}), LHC with $\sqrt{s}=8$~TeV
({\tt COLLIDER=1} and {\tt SQRTS=8000}), and MSTW2008
({\tt PDFSET=MSTW08}). There are four additional predefined PDF sets:
ABM11, NNPDF2.1, JR09~\cite{JimenezDelgado:2009tv}, 
and CT10. For these sets the computation of the
PDF$+\alpha_s$ error is completely automated and it is possible to vary the
value of $\alpha_s(M_Z)$ 
(except for JR09 where no fits with
different values of $\alpha_s(M_Z)$ are provided). Other sets can be
used by specifying the grid file name. It is also possible to choose a
particular member of a PDF set to be used in the calculation. This can
be used to vary $\alpha_s$ with a user specified grid file or for a
re-weighting computation as in Section~\ref{sec:gluon}.

The theory error is determined by the variation of the various scales
(hard, soft, factorization, resummation, and Coulomb) and other
parameters like $\beta_{\text{cut}}$ (cf. \cite{Beneke:2011mq} for a detailed
explanation). The central values of these scales can be chosen by the
user. The variation is done automatically by evaluating the cross
section at several equidistant points between twice and half the
central value. The number of points can also be chosen by the
user. Alternatively, the user can vary the scales by hand by
performing the calculation without error computation ({\tt NOERR=1})
for different values of the scales. This also makes it possible to
evaluate the cross section for specifically chosen points.

The computation time strongly depends on the choice of the PDF set,
with ABM11 being the fastest and NNPDF2.1 the slowest (this was also
noted in \cite{Alekhin:2012ig}). On an AMD Phenom 9850 processor with
2.5~GHz, the computation of a single cross section point (without any 
error estimates, using the
default settings) takes approximately 15 seconds for ABM11, 75 seconds
for MSTW2008, and 6 minutes for NNPDF2.1. Using the default
configuration file to perform the computation of the
NNLO$_{{\rm app}}$ and NNLL$_2$ approximations with MSTW2008, including
the full theory and PDF$+\alpha_s$ error, requires the evaluation of
several hundreds of cross section points and takes about ten hours.

%%%%%%%%%%%%%%%%%%%%%%%%%%%%%%%%%%%%%%%%%%%%%%%%%%%%%%%%%%%%%%%%%%%%%%%%%%%%%

\providecommand{\href}[2]{#2}\begingroup\raggedright\endgroup

\end{document}